\documentclass[manuscript,screen,acmsmall]{acmart}

\AtBeginDocument{%
  \providecommand\BibTeX{{%
    \normalfont B\kern-0.5em{\scshape i\kern-0.25em b}\kern-0.8em\TeX}}}

\setcopyright{acmcopyright}
\copyrightyear{2018}
\acmYear{2018}
\acmDOI{XXXXXXX.XXXXXXX}

\acmConference[Conference acronym 'XX]{Make sure to enter the correct
  conference title from your rights confirmation emai}{June 03--05,
  2018}{Woodstock, NY}
\acmPrice{15.00}
\acmISBN{978-1-4503-XXXX-X/18/06}

\usepackage{subcaption}

\newcommand{\DESIGN}{\textit{DRackSim}}

\begin{document}

\title{DRackSim: Simulator for Rack-Scale Disaggregated Memory Systems}

\author{Amit Puri}
\author{John Jose}
\author{Tamarapalli Venkatesh}
\affiliation{%
  \institution{Indian Institute of Technology Guwahati}
  \state{Assam}
  \country{INDIA}
}
\author{Vijaykrishnan Narayana}
\affiliation{%
  \institution{The Pennsylvania State University}
  \state{Pennsylvania}
  \country{USA}
}

\renewcommand{\shortauthors}{Amit Puri, et al.}

\begin{abstract}
Memory disaggregation has emerged as an alternative to traditional server architecture in data centers that targets improved memory utilization and higher scalability. Disaggregated memory systems involve multiple independent compute and memory pools that are expected to get dedicated hardware support for memory expansion through high-speed cache-coherent interconnects such as CXL. However, CXL memory access is slower compared to local memory, which can significantly impact the system's performance, and requires a series of optimizations to make suitable use of expandable memory. Research in disaggregated systems generates the need for a simulator to evaluate new designs in the practical configurations of hardware disaggregated memory systems.

This paper introduces DRackSim, a simulation infrastructure to model scalable hardware disaggregated memory systems. It models multiple compute nodes, memory pools, and a common interconnect for coherent memory access. An application-level simulation approach simulates an out-of-order x86 multi-core processor with a multi-level cache hierarchy at compute nodes. DRackSim uses a queue-based simulation to model the network interface at the end nodes and the central switch, which simulates remote memory access at both cache block and page granularity. DRackSim also models a centralized memory manager to manage address space in the memory pools. We integrate community-accepted DRAMSim2 to perform memory simulation at local and remote memory locations by using multiple DRAMSim2 instances. Finally, an incremental approach is followed to validate the core and cache subsystems of DRackSim against Gem5. Further, we model various use-case scenarios for disaggregated memory systems and evaluate their performance over various HPC and cloud benchmarks. We rigorously evaluate DRackSim for a wide range of configurations that simulate real-world deployment models and their impact on system performance.
\end{abstract}



\received{20 February 2007}
\received[revised]{12 March 2009}
\received[accepted]{5 June 2009}

\maketitle

\section{Introduction}
\begin{figure}[t]
\centerline{\includegraphics[width=7cm,height=3.5cm]{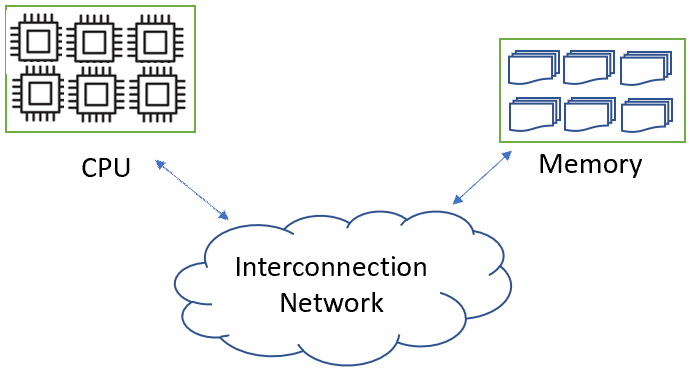}}
\caption{Hardware Disaggregated Memory System}
\label{fig0}
\end{figure}

Data center workloads are increasingly becoming larger in their memory footprints, and traditional server architectures with fixed memory resources no longer remain sustainable for memory resource allocation. With increasing core count, current memory technologies have already reached their scaling limits for bandwidth requirements. Further, the allocation based on peak memory requirements introduces memory under-utilization, and other workloads running on the same server cannot get enough resources to fulfill their memory demands \cite{10.1145/1555754.1555789}. Due to this, a significant imbalance exists in memory usage across servers, while small fragments of memory get stranded within each server node. To overcome this, researchers explored software/virtual disaggregation in the past, which supports free memory at server nodes to be shared among other servers using RDMA (Remote direct memory access) operations over a supported network \cite{10.5555/3489146.3489204,10.5555/2616448.2616486}. The pages are swapped out to remote memory rather than a slow disk to realize the in-memory processing of the workloads. However, software disaggregation involves complex scheduling or load-balancing operations that require large data movement across servers \cite{10.5555/3154630.3154683,10.5555/3489146.3489204} and allows remote memory access only at the page granularity, which is a latency-intensive task. Hardware memory disaggregation is an emerging approach that allows on-demand memory allocation to compute nodes from large capacity on-network memory pools, as shown in Fig.\ref{fig0}. The compute nodes (servers) primarily rely on remote memory for their memory requirements but also have a small amount of local onboard memory. The local and remote memory address space is organized in a flat-address structure at compute nodes, where the remote memory can be accessed through a memory-centric interconnect integrated on-chip at cache block granularity on a last-level cache (LLC) miss. Such interconnects have been proposed earlier for low latency and high bandwidth access to remote memory \cite{7753261,10.1145/2541940.2541965,10.1145/3310360}. CXL 3.0 \cite{Express,9982424} presents similar memory binding fabric specifications for coherent access to pooled memory systems using CXL.memory protocol. It also includes a CXL switch supporting multiple hosts and CXL devices through the CXL (PCIe 6.0) interface and memory-semantic protocols for cache-based access to remote memory. The host interface holds the memory access logic controller and is integrated into the on-chip interconnect (OCI) bus. 

Considering the lack of commercial availability of disaggregated memory systems, a simulation framework is required to translate research ideas into working models by getting enough insight into the system's performance and trade-offs. The focus of disaggregated memory is towards cloud data centers and high-performance computing (HPC) facilities where the disaggregated systems are expected to be configured  at rack-scale with multiple compute and memory nodes. Even though rack-scale scalability is not possible presently, it is possible to configure
\cite{10.1145/3477132.3483561,9252003,10.1145/3339985.3358496,7459397,8342174,Taylor:15,10074901}. A rack-scale disaggregated simulation primarily requires modeling compute nodes, memory pools, and other components like memory managers and an interconnect. An interconnect ties the compute nodes and remote memory pools to allow cache-based remote access to remote memory through network packets.

This paper introduces a simulation framework that models an environment similar to rack-scale memory disaggregation with all the required components mentioned above. \DESIGN{} follows an application-level simulation approach that uses Intel's PIN platform \cite{10.1145/1064978.1065034} and introduces two modes of simulation: a trace-based and a cycle-level simulation model. The trace-based simulation uses main memory access traces with approximate timing information generated through a pintool with an implicit simulation of a multi-level cache hierarchy. Even though memory trace simulation is fast and easily scalable, trace-based simulations usually lack the modeling details and restrict design space exploration. Therefore we also build a cycle-level simulation model that rely on an execution-driven approach  for which an instruction stream is produced by instrumenting a workload with pintool. The instructions execution is then simulated on a detailed x86-based out-of-order simulation at compute nodes. Both simulation modes support a full spectrum of single to multi-core processor architecture and a multilevel cache hierarchy. \DESIGN{} also models all the necessary system-level components to explore the system research space for disaggregated memory systems and provides an opportunity to evaluate new designs rapidly. The compute node model includes a memory-management unit (MMU) for address translation and an address space management unit similar to an OS memory manager with 4-level page tables for allocating memory pages at local or remote memory. The interconnect is based on a queue simulation model and can be configured to meet the bandwidth and latency of the target interconnect hardware by mapping its network parameters. We modify and integrate an open-source cycle-accurate memory simulator, DRAMSim2 \cite{10.1109/L-CA.2011.4}, for simulating DRAM at compute nodes locally and at remote memory pools. The simulator is designed from the top down to simulate multiple compute and memory nodes where a global clock maintains the time ordering of global events such as network access and remote memory access. It uses a multi-threaded approach (one for each compute node) to perform fast and scalable simulations even with many multi-core nodes and memory pools. As real hardware disaggregated memory systems are still in the prototype stage, we use gem5 to perform incremental validation of different components in the simulation framework and perform rigorous testing for the reproducibility of results and portability. The main contribution of our work is as follows:
\begin{itemize}
\item We introduce \DESIGN{}, an application-level simulation framework for rack-scale disaggregated memory systems that can model multiple compute nodes and memory pools with necessary memory management and interconnect simulation. 
\item We present two simulation modes in \DESIGN{} with different levels of details that can be used wherever appropriate.
\item We perform incremental validation of all the components over a range of single and multi-threaded benchmarks. Finally, we compare the performance of large in-memory workloads on \DESIGN{} over various configurations and use cases to show the impact of memory disaggregation and slowdowns due to congestion and contention.
\end{itemize}
The rest of the paper is organized as follows: In the next section, we discuss the prior related work for simulating such systems with their limitations. Next, we discuss the disaggregated memory systems based on which we model different components of \DESIGN{}. Section-4 discusses the design and operations of \DESIGN{}. We discuss the validation aspect in section-5 and use case experiments over vast configurations in section-6.
\section{Related Work}
The first question that arises while building a new simulator is: why yet another simulator? Hardware memory disaggregation is a relatively new research area and an emerging memory architecture. Although software disaggregation and its real-world implementations \cite{10.5555/2616448.2616486,10.5555/3154630.3154683,10.5555/3026877.3026897} have been there for a while now. These systems differ from hardware disaggregation and only support page access to remote memory. The concept of remote memory pools (or memory blades/ memory nodes) with coherent access over memory-semantic fabrics is new and yet to be commercialized. Some simulation/emulation environments exist for hardware memory disaggregation, but all are limited to evaluating only a single node. Lim et al. emulated hardware memory disaggregation on top of the XEN hypervisor \cite{6168955} by marking some allocated memory pages as remote in VMM page tables while adding fixed network latency on memory access. However, emulation platforms cannot predict true memory latency, as local and remote accesses both use the same physical memory.  
On the other hand, some hardware prototypes were also built using FPGA's \cite{7459397,8342174,10.1145/3310360,7753261}, but evaluation on scale is an issue. Some proprietary disaggregated memory implementations exist, such as Intel RSD \cite{intelIntelRack}, Facebook's Open Compute architecture \cite{Taylor:15}, etc, for which not much details are available. Modifying open-source architectural simulators such as Gem5, MARSSx86, Sniper, etc., requires a vast effort to simulate disaggregated memory models while considering their inability to model multiple compute nodes. Further, it will also require the modeling of remote memory managers for remote address space and significant modifications to existing memory managers in compute nodes.  In Daemon \cite{10.1145/3579445}, authors heavily modify Gem5 to model memory disaggregation but test a scaled-up environment with artificial network traffic. Also, the code is not made publically available. Hence, we put extra effort into building a dedicated simulator with good enough accuracy, as it is necessary to model disaggregated memory at a large scale to study various factors impacting performance. With more compute nodes, memory access traffic will generate congestion on the network and contention in queues at remote memory pools. Memory access latency and bandwidth are critical for application performance, and both network congestion and memory contention are significant for performance evaluation. It will also allow for a deeper study of memory management in a scaled-up environment. To our knowledge, no other simulator models such requirements, which is the primary reason behind proposing \DESIGN{}.
\begin{figure}[t]
\centerline{\includegraphics[width=13cm,height=4.75cm]{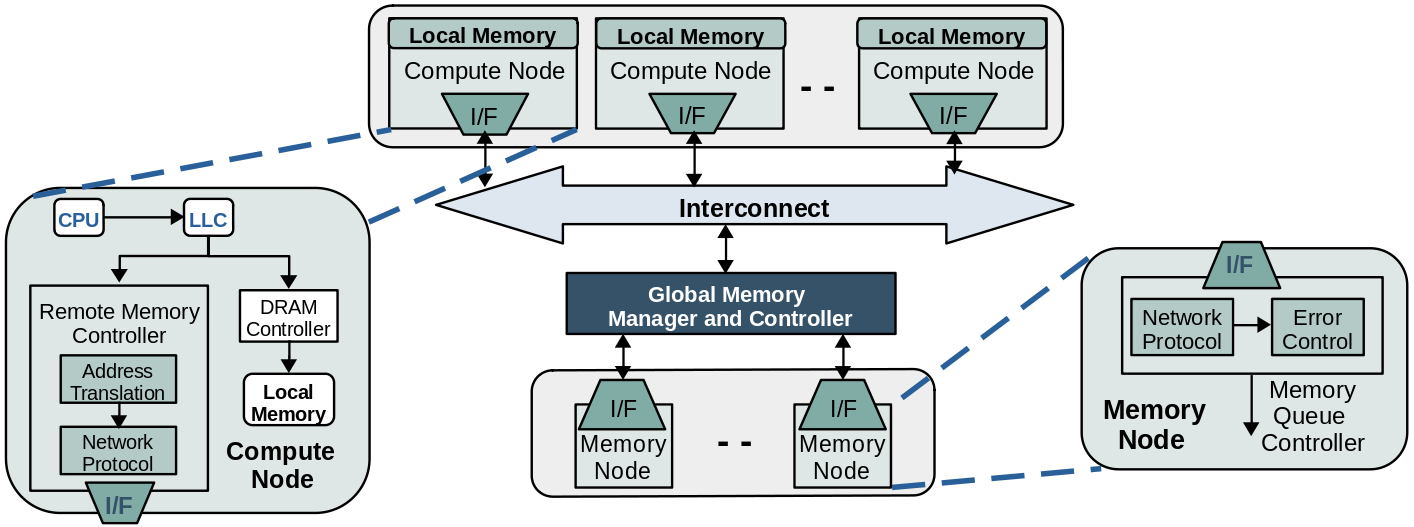}}
\caption{Overview of Disaggregated Memory System and its Interface with Host Compute Nodes; "IF" Interface}
\label{fig1}
\end{figure}

\section{Hardware Disaggregated Memory Systems}\label{3}
Fig.\ref{fig1} shows the abstract view of a hardware disaggregated memory system consisting of compute nodes, memory nodes, interconnect, and a global memory manager/controller. \textit{Compute nodes} are the primary focus for performance evaluation, have a small amount of local memory, and rely on remote memory for most application requirements. \textit{Remote memory pools} are the memory-only nodes (or memory blades) from which a central manager allocates memory to compute nodes on demand. \textit{Global memory manager} is an in-network centralized remote memory manager and controller that performs memory-related activities such as memory allocation, revocation, etc., in the remote memory address space. An \textit{Interconnect} is the binding fabric for supporting memory access by the compute nodes in remote address space. The compute nodes are interfaced to the network interconnect through a remote memory controller, an addressable hardware module similar to a local DRAM controller. The last-level cache misses belonging to the remote memory are forwarded to the remote memory controller, which then performs address translation and implements network protocol in hardware before sending it to the physical layer interface (root port in CXL). A similar controller is present at the memory nodes to decode the network memory request packets and send back the responses to CPU nodes.

\begin{figure*}[t]
\subcaptionbox{\label{2b}}
{
{\includegraphics[width=3cm,height=3.25cm]{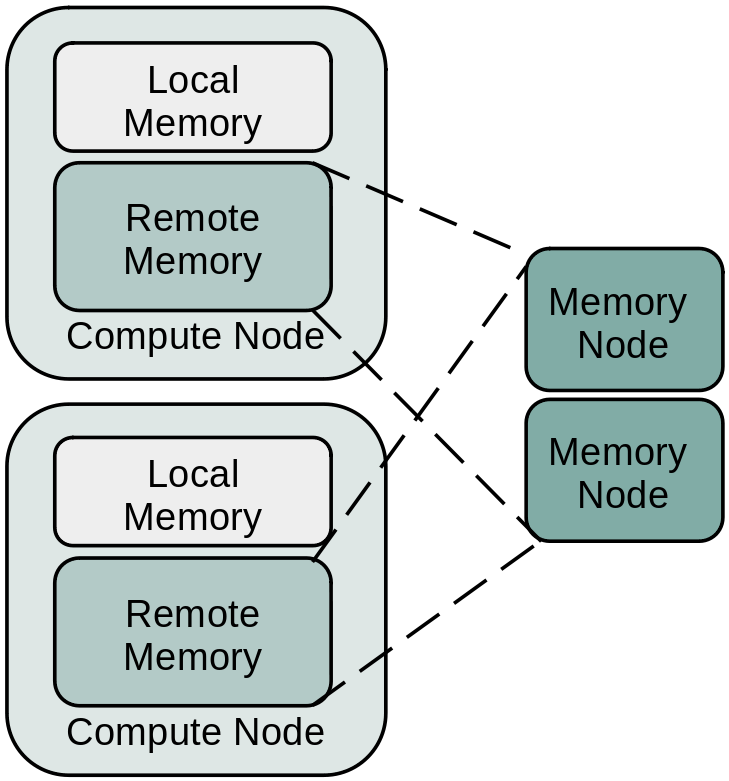}}
}
\hfil
\subcaptionbox{\label{2c}}
{
{\includegraphics[width=3cm,height=3.25cm]{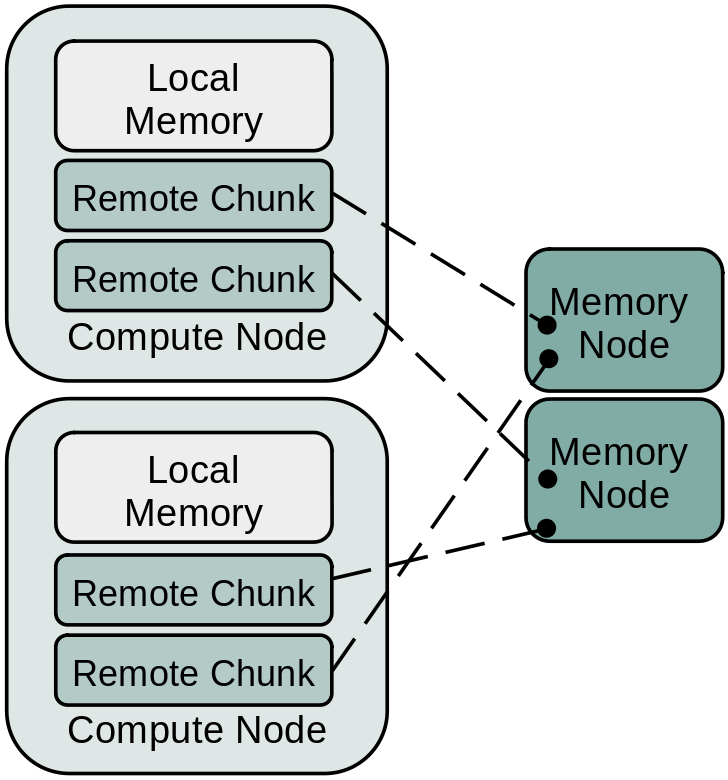}}
}
\hfil
\subcaptionbox{\label{2d}}
{
{\includegraphics[width=7cm,height=3.75cm]{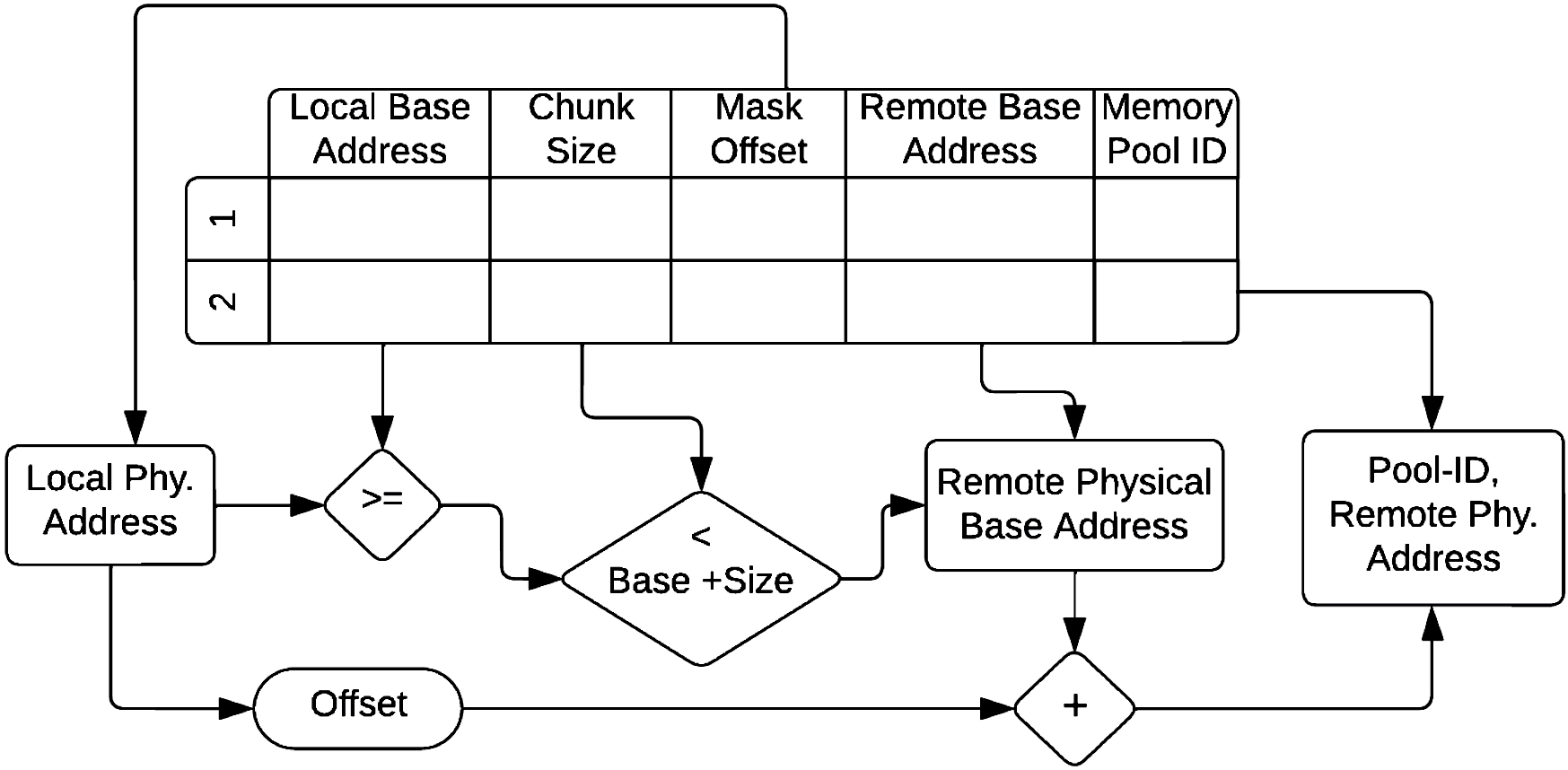}}
}
\caption{Remote Memory Exposure to Compute Nodes (a) Shared Organization (b) Distributed Organization (c) Address Translation at Remote Memory Controller}
\label{fig2}
\end{figure*}

\subsection{Remote Memory Organization}
The scalability in hardware disaggregated memory will largely depend on the remote memory organization schemes and the way remote memory is exposed to the system at compute nodes. Remote memory can be organized using a shared memory or as a distributed approach, as shown in Fig.\ref{2b} and \ref{2c}, respectively. In the shared memory approach, all the remote memory address space is visible to the OS at compute nodes with a single global address. The application can be allocated a page at any address while supporting page sharing between nodes, with the owner node acting as a home agent for that page. Such memory organization will still generate excessive memory coherence traffic (to memory nodes and other compute nodes) in the network due to shared pages, limiting the system's scalability. The compute nodes will also require frequent communication with a central authority (global memory manager) to prevent an address conflict during a remote page allocation, creating a bottleneck in the global memory manager while serving page allocation requests. But the advantages are that the applications can also span across multiple servers to meet the computing requirements. 

However, unlike software-disaggregated systems, which rely on memory sharing across the servers for scalability, hardware memory disaggregation does not need to do so for two reasons. Firstly, scalability can be achieved through sufficiently large pools of memory, and the compute nodes can directly request remote memory from the memory nodes whenever required. Secondly, high-end multi-processor systems support tens of cores, and most data-center applications can fulfill their computing requirements within a single node. Even if the workflows get spanned across multiple nodes, they run independently with occasional communication. The distributed memory organization can fulfill such requirements. In this organization, remote memory address space is not initially visible to compute nodes. The remote memory can be reserved in chunks of size, say 4MB-16MB (easy to evict and deallocate chunks), whenever a node requests. This approach will also require a global memory manager to reserve remote memory for a compute node, but allocation in larger chunks will not create a bottleneck. Considering these benefits, \DESIGN{} models distributed approach and uses an extra layer for address mapping at compute node's remote memory controller, as shown in Fig.\ref{2d}.  This mapping is required for translating the local physical addresses to the remote physical address for the allocated remote memory chunks, which differs from virtual to physical address translation at TLBs. Frequently used addresses can be kept in the cache at the remote memory controller, which will add a few extra cycles on each remote memory access. The Linux memory hot-plug and hot-unplug service supports adding or removing memory to the compute node's address space during run-time. Once initialized, the new memory is available as an extension of the local address space that can be used for regular page allocation. 

With the distributed approach, compute node will have exclusive access to remote memory chunks allocated to it. The coherency traffic is limited between compute and memory nodes, and the need to maintain inter-compute node coherence is eliminated while enhancing scalability. Further, the major part of compute node's cache capacity will consist of remote memory data, and the types of caches (write allocate/no-allocate, write-through/write-back) is a crucial factor for the extent of coherence traffic between compute and memory nodes. The outward coherency traffic to memory pools can be minimized using write-back caches at compute nodes.
\begin{figure}[t]
\centerline
{\includegraphics[width=14cm,height=4.5cm]{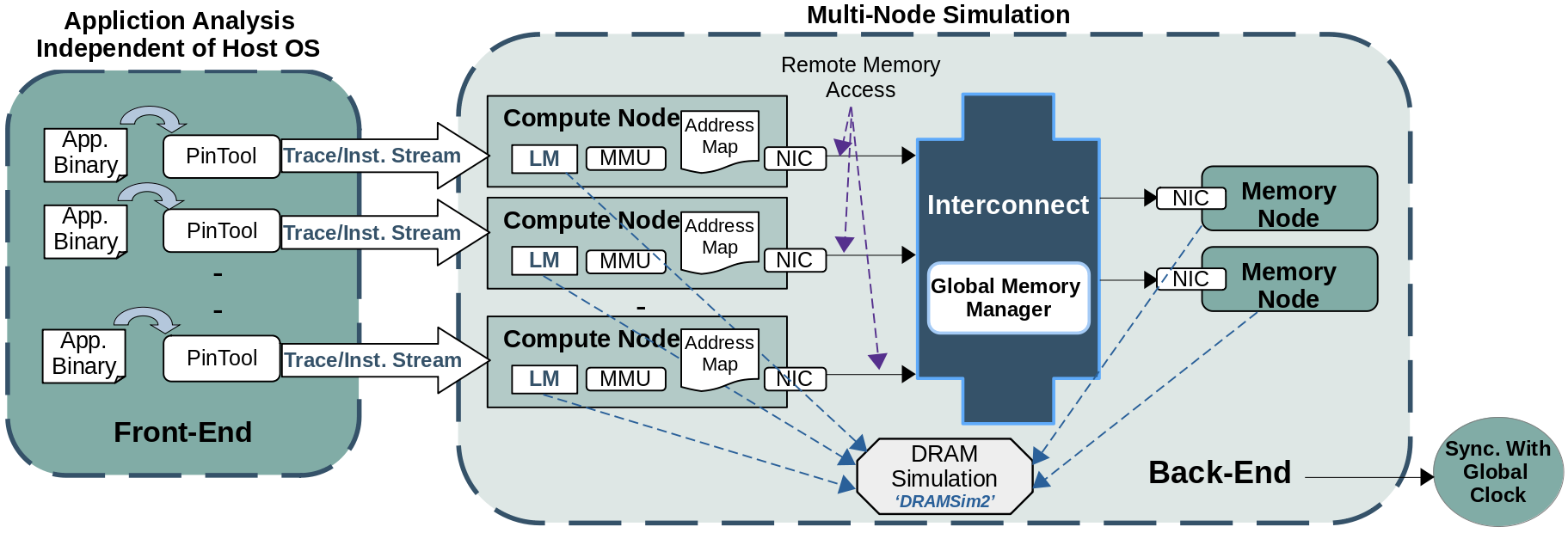}}
\caption{\DESIGN{} infrastructure overview; "LM" Local memory, "MMU" Memory management unit, "NIC" Network Interface}
\label{fig4}
\end{figure}

\section{DRackSim Design and Operations}
The objective of \DESIGN{} is to enable researchers to explore new hardware structures (efficient page migrations or prefetching mechanisms) and system-level designs (memory management at compute nodes and globally) for disaggregated memory systems. Therefore, we carefully model the required components to gain enough insights for performance evaluation. Fig.\ref{fig4} gives an overview of the complete simulation process in \DESIGN{} with its two-phase design: a front-end and a back-end. A Pin-based front-end performs the application analysis, whose output is fed to the back end for scalable multi-node simulation with disaggregated memory. \DESIGN{} supports two different modes of operation: a trace-based simulation model and a cycle-level detailed simulation model.

\subsection{Trace-Based Model}
\begin{figure}[t]
\centering
\subcaptionbox{\label{5a}}
{
\includegraphics[width=4.5cm,height=4cm]{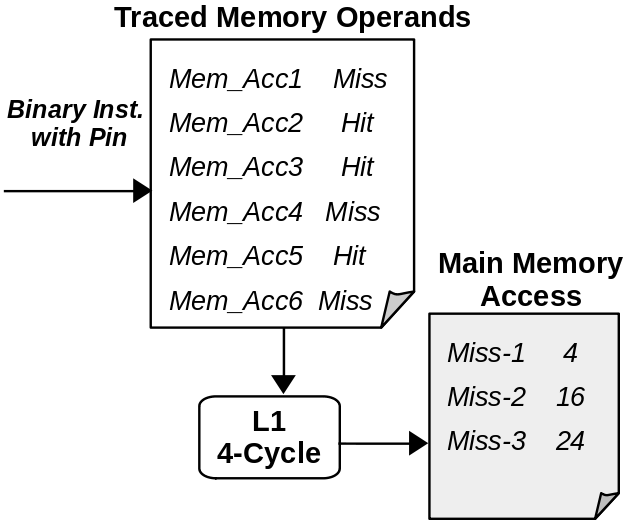}
}
\hfill
\subcaptionbox{\label{5b}}
{
\includegraphics[width=3cm,height=4cm]{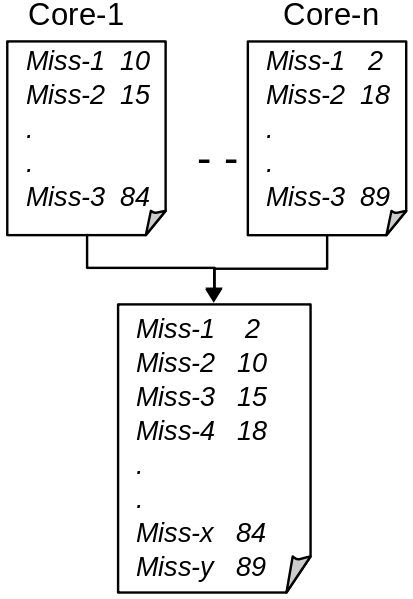}
}
\hfill
\subcaptionbox{\label{5c}}
{
\includegraphics[width=5cm,height=5.5cm]{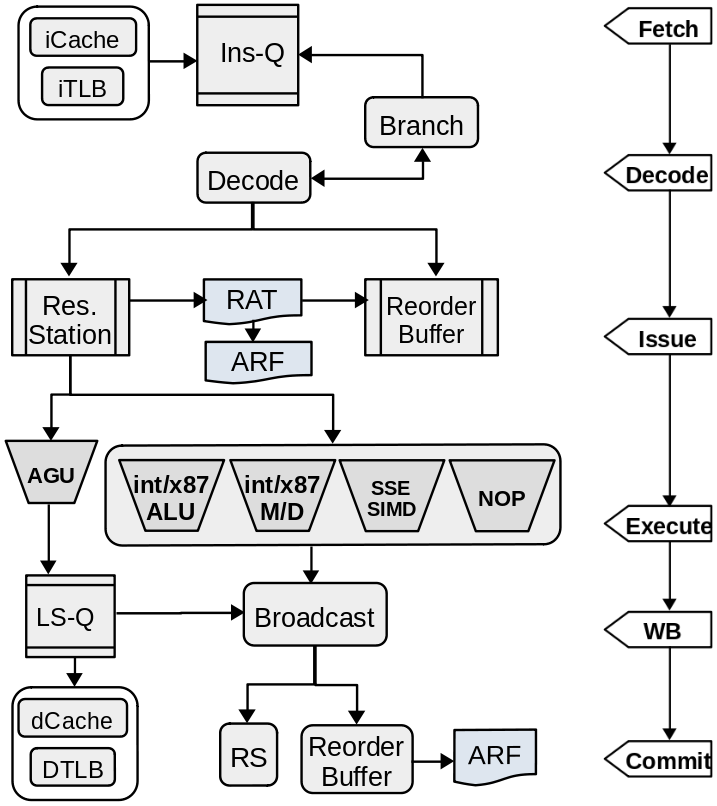}
}
\caption{Trace Generation (a) Recording Main-memory access (b) Final Multi-Threaded Trace ; Cycle-level Simulation (c) Out-of-Order Core Modeling Subsystem}
\label{fig5}
\end{figure}

The trace-based simulation is fast but less accurate and can quickly evaluate disaggregated systems on a large scale for billions of local and remote memory accesses. Pin provides with its package a tool named \textit{Allcache} that performs a functional simulation of the single-core cache hierarchy. We extend this tool to support multi-core TLB and a 3-level cache hierarchy (private I/D-L1, L2, and shared L3) and add support to instrument multi-threaded workloads. The instrumentation is done at the instruction-level granularity, and each thread is mapped to one of the cores based on its thread ID. The tool generates memory references whenever an instrumented instruction has a memory operand. The memory references are passed through the TLB/cache model to generate an approximate cycle number for each LLC miss.
With the help of access latency at each cache level and the hit/miss status of all the memory accesses, an aggregate counter is maintained to determine the clock cycle number. If the memory access is a miss at LLC, it is recorded in a trace file with the aggregate counter. Fig.\ref{5a} shows an example of generating main memory traces on a single cache level with 4-cycle latency. A similar process is followed to collect LLC misses at different cores which are sorted and merged together to generate a single trace file representing all the main memory accesses while also maintaining the workload's multi-threaded nature, as shown in Fig.\ref{5b}. Even though main memory traces are less accurate than a real CPU model driving the memory model, they are convenient to use and consume significantly less disk space than CPU-generated memory traces. However, traces are static and do not allow system-level optimizations like hot-page migration, page-swapping systems, cache prefetching, etc., that change the state of TLBs and caches during the simulation.

\subsection{Cycle-Level Simulation Model}
Due to some known limitations of the trace-based model, we present a cycle-level simulation model for the multi-core out-of-order x86 processor architecture at the compute node. In this mode, the pintool scope is restricted to producing an instruction stream by intercepting each executed instruction in the workload. The trace consists of instruction type (int/x87 ALU, int/x87 mul/div, SSE (vector), branch, nop, etc.), instruction address, memory operand address/size, and register dependencies for each instruction. The core and cache subsystem modeling is handled at the back end in this simulation mode.

\subsubsection{Out-of-Order Core Modeling}
Fig.\ref{5c} shows the details of the OOO pipeline core modeling subsystem. The core architecture implements multiple pipeline stages (fetch, decode, issue, execute, write-back, and commit) at a higher level of abstraction with detailed modeling of hardware structures such as instruction queue, reservation stations, re-order buffer, architecture register file (ARF), register-alias table (RAT), and load-store queue. The instructions are read from the instruction stream generated by Pin, after which a fetch unit simulates the instruction fetch for multiple instructions within a cache line rather than fetching them individually. This is also pointed out in the previous work as many other academic simulators fetch each instruction separately \cite{6557148,4919638}. The decode unit decodes the instruction, puts it into a buffer, and checks for the branch and its prediction result. The instruction waits in the decode buffer until the hardware resources are allocated. A limitation of performing binary instrumentation is that the execution of a process is decoupled, and it never goes down the wrong branch path in simulation. But the Pin API can tell whether an instruction is a branch. A branch predictor matches the prediction result with the information passed by the pin, and a penalty is added in case of misprediction, during which the CPU remains stalled. The issue unit allocates an entry for the instruction in the reservation station (RS) and re-order buffer (ROB) or stalls it if no free entry is available. The incoming instruction in the RS clears its register dependencies by accessing the register file (ARF) or from the register alias table (RAT), which points to a ROB entry. The instruction waits if some previous instruction does not yet free a register. The memory read operands (if present) are sent to an address generation unit (AGU) to simulate effective address calculation and forward the address to the load-store queue for memory access. If the same load address is already present in the queue as a store, it is forwarded without waiting. Once the dependencies are clear, instructions are moved to a ready queue. The dispatch unit selects and allocates execution units based on the instruction type and opcode. The execution latency can simply be configured for each type of instruction and its operation based on the number of cycles it takes to execute in the target processor model. Finally, the result gets broadcast among all the hardware structures: the waiting RS entries clear their memory or register dependency, instruction status changes to 'executed' in ROB, and a write-back is performed to memory if there is a write operand. Only in the commit stage is the ROB entry released, and updates to the register file are performed to make it available globally. \DESIGN{} allows the user to conFig.all the CPU parameters to simulate target hardware.

The cache model comprises a multi-level hierarchy with private L1 I-D, L2, and shared L3 cache. The non-blocking caches support multiple outstanding misses using miss-status handling registers (MSHRs) with a configurable number of entries. The memory access for instruction fetch or load/store queue starts at the TLB for virtual to physical address translation and uses 4KB fixed-size pages. Once the memory access reaches LLC MSHR, it is queued for the main memory access and DRAM simulation. The caches can be configured to be either write-allocate or no-write-allocate. Finally, the cache subsystem notifies the corresponding entry in the load/store queue on completing a memory request, which is then broadcast to the waiting instructions in the pipeline.

\subsection{Back-end Modeling}
The back-end of \DESIGN{} simulates an environment similar to a large-scale disaggregated memory system with multiple compute nodes running simultaneously on different simulation threads. The memory accesses produced by the compute nodes drive the DRAM simulation at local or remote memory. The local memory requests are directly simulated at the node's local memory, and remote memory requests are passed through an interconnect model before being simulated for memory access at the remote memory pool. We explain here all the simulated components to model disaggregated memory behavior.

\subsubsection{Compute Nodes}
Besides CPU and cache simulation, compute nodes models a local memory unit and a memory manager to make memory allocation decisions and manage address space. The memory manager is an abstraction of processor MMU for address translation and an OS-like memory manager for address space management and memory allocation at the compute node. A memory request reaches MMU on a TLB miss and performs a page-table walk with a defined latency. If the page is not in memory, the request is forwarded to the page-fault handler for memory allocation and creates a page-table entry (PTE). \DESIGN{} models 4-level page tables for mapping virtual addresses to the physical pages. The memory manager allocates a new page in local or remote memory based on the allocation policy and availability of free memory space. The page-fault service stalls the CPU and incurs fixed latency, which can be configured to model the OS page-fault latency. The page allocation in disaggregated memory is crucial, and the memory manager should carefully decide the footprint ratio in local and remote memory for different applications running at compute node. The response time of latency-sensitive applications can be significantly impacted compared to a less latency-bound application, and an uninformed page allocation policy can be responsible for high tail latency. 
The memory allocation at compute node in \DESIGN{} can be configured to allocate memory pages in any ratio from local and remote memory. The modeling of these components allows for exploring memory management policies that can make such decisions. 

\subsubsection{Global Memory Manager}
The global memory manager (and controller) takes care of the remote memory address space in all the memory pools and reserves remote memory from one of the pools on receiving a request from compute nodes. The global manager handles conflicts during remote memory reservations to different nodes and acts as a load balancer while choosing a memory pool for allocation. On reservation of a memory chunk, it will share chunk details (pool-id, remote base address, size, etc.) with the requesting compute node, creating an entry in its mapping table as shown in section \ref{3}. The memory pools are bound to face contention in their queues when several compute nodes access the same pool simultaneously. To avoid contention and tail latency, memory pool selection should be done so that all pools face almost similar amounts of memory request traffic. It should also ensure Quality of Service (QOS) to different applications running on compute nodes with different sensitivities towards memory latency (compute vs memory bound applications) using request scheduling mechanisms. \DESIGN{} follows a round-robin pool selection while reserving a memory chunk and allows further exploration of similar pool selection policies.

\subsubsection{Interconnect Model}
The interconnect model in \DESIGN{} is based on a queue simulation that simulates the behavior of memory-semantic fabrics such as CXL or other similar interconnects proposed for disaggregated memory. The interface (similar to NIC) at the compute nodes (memory requestor) and memory pools (memory responder) allows access to remote memory through the network. The on-chip integration of the fabric and lightweight network protocol implementation in the hardware allows low-latency cache line access from remote memory during an LLC miss. The CXL specification allows cache-based access to remote memory from compute nodes with a latency of around 170-250ns. Similarly, a 4KB remote page (64 cache lines of 64B each) can be accessed in around 1.2-1.5µs. The remote memory accesses from multiple compute nodes pass through a central switch before accessing the pooled memory.

\begin{figure}[t]
\centering
\subcaptionbox{\label{7a}}
{
\includegraphics[width=9cm,height=3.25cm]{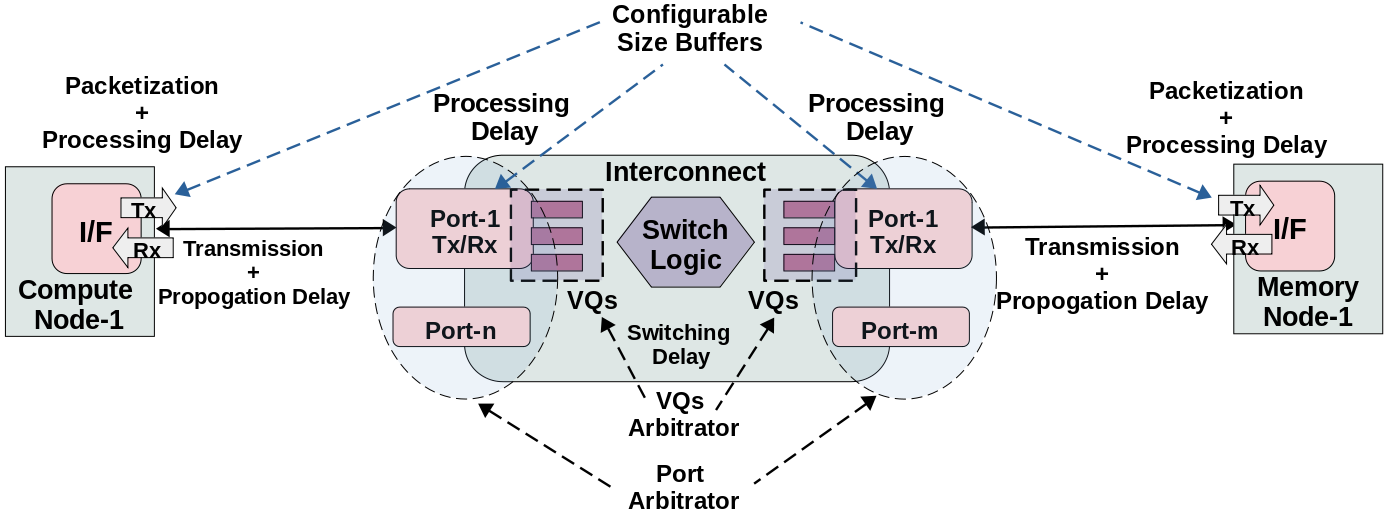}
}
\hfil
\subcaptionbox{\label{7b}}
{
\includegraphics[width=3.75cm,height=2.5cm]{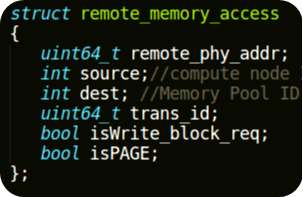}
}
\caption{(a) Interconnect Simulation Model detail (b) Packet Structure for Remote Memory Access}
\label{fig7}
\end{figure}

\DESIGN{} simulates both types of memory accesses from the compute nodes to remote memory pools for design space exploration. If an LLC miss refers to remote memory address space, it accesses the local-to-remote address map (discussed above) at the compute node's remote memory controller. The memory access is encapsulated into a network packet containing the destination memory pool-id and its remote physical address, as shown in Fig.\ref{7b}. The model uses fixed 64-byte packets for a memory request, as the payload consists of only a memory address. The packet is then pushed into the queue structure at the controller's network interface after adding a delay for packetization. While the packet transmits from compute node to the switch, it incurs transmission and propagation delays based on configured bandwidths at the nodes and hop distance. It is then added to the switch input port queue and faces a processing delay. The interconnect model implements a crossbar topology that supports single-hop communication between compute and memory nodes. It supports virtual queues at switch ports to avoid head-of-line blocking, and a 2-stage switch arbitrator selects the packet for forwarding in each cycle. The first stage arbitrator selects one of the input ports, and the second stage selects from one of the virtual queues at the selected input port. The packet is added to the buffer at the destination output port after adding a switching delay. Finally, it reaches the network interface of the destination memory pool to simulate the remote memory access. A similar way is followed at the memory pool to send back the response to a compute node using the source address of the memory access packet. The response packet holds a cache line of data as a payload for block accesses, and its size can be configured accordingly based on cache block size. Similarly, write-backs from compute nodes to remote memory also use a packet size capable of storing a cache line. 

Further, page requests are also simulated from compute nodes to remote memory. Memory pools can differentiate between block or page request packets through the additional information present in the header. The response from the memory pool can be sent as single or multiple small-sized packets that the user can configure. The reassembly logic collects all the response packets at compute node to form a memory page and notify on receiving a complete page. This functionality can be used to implement hot-page migration systems to reduce average memory latency. However, poorly scheduled page access can starve the critical block accesses and should only be used to supplement cache line accesses for utilizing locality in hot pages. Further, the network interface and the ports at the switch support configurable size buffers at both ends and implements a reliable network with back-pressure flow control in case a buffer gets full. The interconnect model latency of \DESIGN{} can be mapped to simulate target hardware (CXL/GenZ) for cache line and page transfer. Fig.\ref{fig7} shows a complete view of interconnect simulation model in \DESIGN{}.

\subsubsection{Memory Simulation}
We integrate cycle-accurate DRAMSim2 for DDR4 simulation of local and remote memory. We initialize multiple instances of DRAMSim2 memory units using its \textit{Memory System} interface, each representing either the local memory at a compute node or remote memory at a memory pool. DRAMSim2 provides a \textit{callback} functionality to notify the CPU driving the memory model on completing every memory access. We modify the \textit{MemorySystem} interface and \textit{callback} functionality so that each memory unit (at a node or remote pool) can have a separate identity. We further modify the \textit{addtransaction} function, which is used to add a memory request to a memory unit, to include a node-id of the requester, a transaction-id, and some other metadata for stats collection. The modifications allow tracking the completion of memory accesses at each memory unit separately and correctly sending back a response to the requesting node.

\subsubsection{Simulator Implementation and Clock Management}
The back end of \DESIGN{} handles the scalable disaggregated memory simulations of multiple compute and memory nodes. In the trace-based model, multiple traces are collected (one for each node) and parsed in parallel to perform memory and network simulation. The memory requests are split across local and remote memory based on their access address and the location of respective pages. In the cycle-level simulation, multiple instruction streams are generated simultaneously by different Pintool instances (one for each node). All the instruction streams are continuously fed to the back end for multi-node disaggregated memory simulation. We also add support for instrumenting multiple applications to create workload mixes for a single compute node. The user can skip any number of instructions to jump to the region of interest in a workload. As the back-end, \DESIGN{} model consists of multiple independent components such as compute nodes, memory pools, interconnect, etc., which are synchronized with a single global clock. This is necessary to maintain the time-ordering of global events, such as simultaneous network and remote memory accesses from different compute nodes. However, the frequencies of individual components can be configured separately, and the global clock only provides a common reference time for the functional correctness of the simulation. We utilize \textit{thread-barriers} for synchronization that controls the simulation flow for multiple nodes (with each node being simulated by a separate simulation thread), allowing scalable simulation without much slowdown.
\section{Validation}\label{val}
It is important to cover different validation aspects while developing a new architectural simulator. The first one is the functional correctness of the program. In our case, application functionality is decoupled from the actual simulation process, where Pin runs it natively. Pintool only adds some instrumentation primitives and does not change the application's functionality or execution flow. The second aspect is the accuracy of performance metrics. While it is important to validate the simulator with actual hardware or with another standard open-source simulator, the vast performance gap between various community-accepted architectural simulators is also a reality.

\begin{table}[htbp]
\caption{Validation Parameters}
\begin{center}
\begin{tabular}{l|l}
\toprule
\textbf{Element} & \textbf{Parameter}\\
\toprule
\toprule
CPU & 3.6GHz, 8-width, 64-InsQ, 64-RS, 192-ROB, 128-LSQ\\
\hline
L1 Cache & 32KB(I/D), 8-Way, 2-Cyc, 64B block\\
\hline
L2 Cache & 256KB, 4-Way, 20-Cyc, 64B block\\
\hline
L3 Cache & 2MB per core shared, 16-Way, 32-Cyc, 64B block\\
\hline
Cache Type & Write-Back/Write-Allocate, Round-Robin\\
\bottomrule
\end{tabular}
\label{tab1}
\end{center}
\end{table}

Due to the unavailability of a full-scale hardware disaggregated system, we incrementally validate different components of our simulator. The integrated on-chip interconnects, such as CXL, are either commercially unavailable or only have been tested with small-scale prototypes using FPGAs \cite{9137193,10.1145/3310360,7753261}. Therefore, it is a best-effort approach to validate the core and cache subsystems of DRackSim against Gem5 system emulation mode (SE). At the same time, we show the impact of network interconnect separately through our wide set of experiments in the next section. We set the processor width and instructions latency for different instruction types to the same values for calibrated validation and used the same size structures for all the hardware resources (such as InsQ,RS,ROB,LSQ). We further fix our simulator's page fault and TLB-miss latency as per Gem5, which only adds 1 or 2 cycles on each such event in SE mode. Table \ref{tab1} shows the system configuration for CPU validation. We extensively validate the cache subsystem using the last-level cache misses, which can represent the behavior of the complete cache hierarchy.

\begin{figure*}[t]
\centering
\subcaptionbox{\label{8a}}
{
\includegraphics[width=6.74cm,height=3cm]{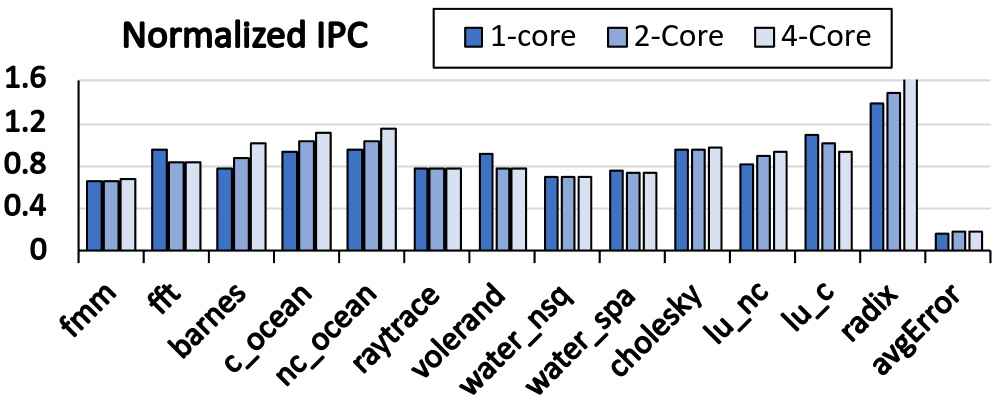}
}
\subcaptionbox{\label{8b}}
{
\includegraphics[width=6.74cm,height=3cm]{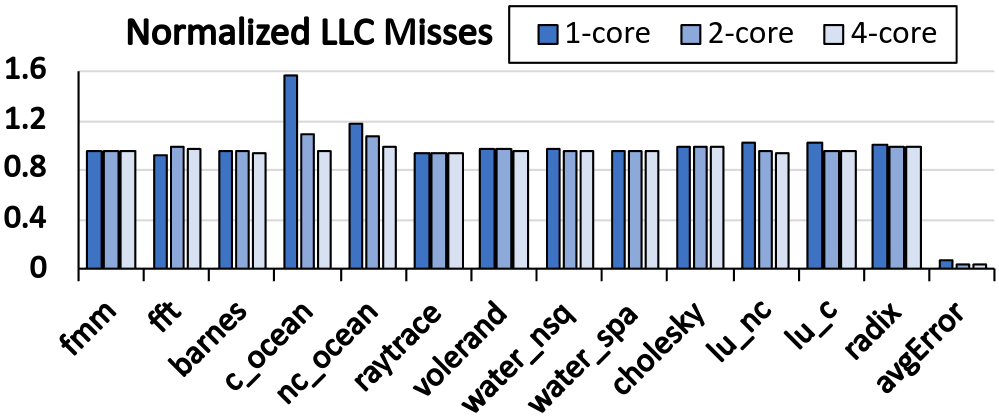}
}
\caption{ Validation on \textit{Splash-3} benchmarks (a) variation in IPC values (b) variation in L3 Cache Misses}
\label{fig8}
\end{figure*}

\begin{figure*}[t]
\centerline{\includegraphics[width=13cm,height=3.5cm]{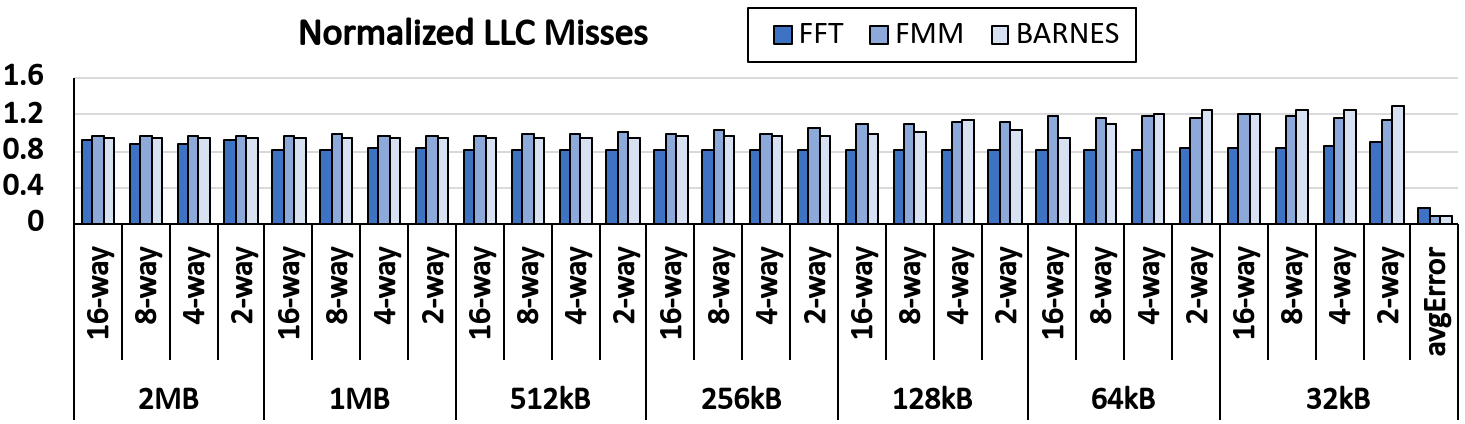}}
\caption{Normalized L3 Cache Misses over different cache configurations}
\label{fig10}
\end{figure*}

We perform the CPU validation for 1, 2, and 4-cores over SPLASH-\cite{7482078} benchmarks by spanning same number of threads as the number of cores. Figure \ref{8a} shows the CPU validation results with normalized IPC values of our simulator compared to the IPC values of Gem5. The IPC numbers of \DESIGN{} are close to Gem5 IPC for most of the benchmarks and show a mean absolute percentage error of 15\% across all core configurations. Similarly, we validated LLC misses for all the workloads, as shown in figure \ref{8b}. We only observe an unexpectedly large error in the case of \textit{contiguous\_ocean} with a 1-core CPU. Besides this, the mean absolute percentage error is around 3\% across all benchmarks on 1, 2, or 4 cores CPU. However, the variation observed is common among different simulators, as shown for the validation efforts in the past work \cite{6557148,8718630}. Ayaz et al. surveyed all the major architectural simulators, such as Gem5, MARSSx86, Sniper, etc., and observed their performance against real x86 hardware. It was found that a significant variation exists between the IPC values and LLC misses on all benchmarks. The main source of inaccuracies could be due to the lack of support for fused micro-operations (\begin{math}\mu ops\end{math}), the pipeline depth, lack of modeling for all hardware optimizations, and the abstraction of actions performed during branch misprediction. However, the simulator can be calibrated to match the performance of real hardware and should provide enough insight into actual hardware performance.

We further perform an in-depth validation for the LLC misses with \textit{fft, fmm, and barnes} over a range of L3 configurations on a 4-core CPU, shown in figure \ref{fig10}. We also reduce the L2 size to 64KB to maximize the misses at LLC. Here also, we observe a slight variation in the LLC misses for DRackSim compared to Gem5. We observe a mean absolute percentage error of 11\% over all the configurations. The LLC misses are slightly inflated or deflated for some configurations, which could be due to the implementation details of the cache hierarchy. We do not implement a separate write buffer, so the caches must evict a block during the write-backs. Another reason can be using separate load and store queues in Gem5, whereas DRackSim has a unified load/store queue that can create a small difference in the total number of non-redundant loads and stores. These differences can generate variation in cache accesses, and inaccuracies can accumulate from lower-level caches to LLC.
\section{Evaluation}
In this section, we demonstrate the working of \DESIGN{} over various configurations and provide use case evaluation by modeling different data movement schemes in disaggregated memory systems. We also show the impact of the network by changing the latency and bandwidth parameters at the nodes interface and the switch.

\begin{table}[htbp]
\caption{Benchmarks}
\begin{center}
\begin{tabular}{llll}
\toprule
\textbf{Application} & \textbf{Domain} & \textbf{Input} & \textbf{LLC Miss}\\
& & & \textbf{(Millions)}\\
\toprule
\toprule
Stream Cluster (SC) \cite{5306797} & Data Mining & Pts:65536 Dim:256 & 3.67\\
\hline
Needleman Wunsch (NW) \cite{5306797} & Bio-informatics & Rows/Col:4096 Pen:4 & 6.68\\
\hline
Block Tri-diagonal (BT) \cite{8374543} & CFD & Class C & 2.97\\
\hline
3D Fast Fourier (FT) \cite{8374543} & CFD & Class C & 28.63\\
\hline
High Perf. Conj. Grad. (HPCG) \cite{hpcg} & HPC & 104 \begin{math} \times \end{math} 104 \begin{math} \times \end{math} 104 & 2.71\\
\hline
K-core Decomposition (KD) \cite{10.1145/2442516.2442530} & Graph Proc. & V:1M E:10M & 0.95\\
\hline
K-means Clustering (KC) \cite{5306797} & Data Mining & kdd\_cup & 0.62\\
\hline
Lulesh (LU) \cite{osti_1090032} & HPC & Cube Mesh Size:120 & 5.29\\
\hline
Multi-Grid (MG) \cite{8374543} & CFD & Class C &  37.95\\
\hline
miniFE (FE) \cite{osti_993908} & HPC & 140 \begin{math} \times \end{math} 140 \begin{math}\times\end{math}140 & 14.89\\
\hline
PageRank (PR) \cite{10.1145/2442516.2442530} & Graph Proc. & V:1M E:10M & 0.95\\
\hline
Particle Filter (PF) \cite{5306797} & HPC & 2K \begin{math} \times \end{math} 2K 20K Particles & 33.13\\
\hline
Pennant (PEN) \cite{https://doi.org/10.1002/cpe.3422} & HPC & leblancx4.pnt & 8.02\\
\hline
SimpleMOC (SM) \cite{Gunow2015} & HPC & small & 1.75\\
\hline
SRAD (SR) \cite{5306797} & Image Proc. & 4K \begin{math} \times \end{math} 4K Data Points & 3.19 \\
\hline
XSBench (XSB) \cite{Tramm:wy} & HPC & small & 8.32\\
\bottomrule
\end{tabular}
\label{tab2}
\end{center}
\end{table}

We evaluate the performance using various multi-threaded benchmarks shown in Table \ref{tab2}. We chose five workloads from the Rodinia heterogeneous benchmark suite: SC, NW, KC, PF, and SR, that simulate applications from different domains such as data mining, bio-informatics, image processing, etc. Two graph processing workloads, KD and PR, were chosen from the Liagra framework. Three workloads, BT, FT, and MG, are selected from the NASA parallel benchmark suite that mimics the computation and data movement in computational fluid dynamics applications. Further, we use six more multi-threaded workloads and mini-apps from the domain of HPC: HPCG, LU, FE, PEN, SM, and XSB. The selected workloads have a wide range of memory footprints, ranging from 38MB to 3.2GB, and vary in memory access patterns. We run openMP versions of all the workloads and only simulate the multi-threaded regions of the workloads (except PF, which has a significantly large single-threaded phase) using 4-threads with each thread spanning one of the cores (no hyper-threading). The simulations were run for at least 400 million instructions for each multi-threaded workload and 100 million for PF, which was slow and took a long time. The last column of the table shows the total number of LLC misses or main memory accesses for all workloads during the simulation period. 

\begin{table}[htbp]
\caption{Simulation Parameters}
\begin{center}
\begin{tabular}{l|l}
\toprule
\textbf{Element} & \textbf{Parameter}\\
\toprule
\toprule
Memory (Loc/Rem) & 2400MHz DRAM (19.2GB/s)\\
\hline
Switch & 100/400Gbps, 4MB port Buffer, 5/10ns for processing and switching\\
\hline
Network Interface & 40/100Gbps, 1MB buffer, 15/30ns for (de)packetization and processing \\
\hline
Packet Size & 64B Request, 128B Response/Write-Backs, 4KB for Page Access\\
\bottomrule
\end{tabular}
\label{tab3}
\end{center}
\end{table}

Finally, table \ref{tab3} shows the network and memory parameters for the simulation, while we use the same CPU and cache parameters mentioned earlier in section \ref{val}. We assume a 64B data packet for the remote memory request packet from the compute node. On the other hand, the response packet from the memory pool and remote memory write-backs will contain a cache block of data and have a packet size of 128B. The remote page request is also 64B, whereas the page access is performed by adding 64 cache block requests to the memory unit, and the response is sent as a single 4KB packet. For interconnect, we consider 100/400 Gbps bandwidth for the switch and 50/100 Gbps bandwidth at node interfaces. We also vary the latency, which is 15 or 30 ns at nodes for (de)packetization and processing and 5 or 10 ns for processing and switching at the switch. Each remote memory access will pass four times through the node interfaces (request/response at compute node and memory pool) and two times through the switch (request/response), bringing the total latency factor to 70 or 140 ns.

\subsection{Design Space Exploration}
To understand the impact of memory disaggregation with traditional server systems, we modeled four use case scenarios with \DESIGN{}: {\large \textcircled{\small 1}} \textbf{Block:} The first is our baseline hardware disaggregated memory system in which all the remote memory accesses are made at the cache block granularity on an LLC miss. The pages are allocated alternatively in local and remote memory, each having 50\% of memory footprint. {\large \textcircled{\small 2}} \textbf{Page:} Next, we model a software disaggregated memory where the remote memory accesses are always made at page granularity as in an RDMA-based memory sharing servers. The page allocation is performed in the same manner with 50\% of the workload footprint at remote memory and is migrated to local memory on the first reference to it. {\large \textcircled{\small 3}} \textbf{Block+Page:} Next, we model a very simplistic page migration on top of the baseline hardware disaggregated system that only migrates some of the hot pages in remote memory to local and utilize locality for future accesses. The memory requests to other remote pages are still made at cache block granularity. The page migration threshold is set through a simple training phase and is fixed at 20\% access count of total accesses to predicted hot pages during training. Further, no special support is added for scheduling page migrations. {\large \textcircled{\small 4}} \textbf{Local:} Finally, we model a local-only system that assumes big enough local memory to fit the entire workload footprint. This system will have no remote page allocation, and all memory accesses are performed in local memory at block granularity.

The systems described in {\large \textcircled{\small 2}} and {\large \textcircled{\small 3}} require an update to system page tables on every remote page migration and invalidate TLB entries on all the cores. This introduces long CPU stalls and may further take around \begin{math}4\mu s - 13\mu s\end{math} (depending on the core count) \cite{10.1145/3364179} for TLB-shootdown after every migration. Therefore, for a fair comparison, we perform page table updates in large batches of 1024 pages and use a remap table to delay page migrations by temporarily storing the new physical address of a migrated page.

We start our evaluation with a single compute node with one memory pool to show the performance impact of all three configurations compared to local. With a single compute node, all the network and remote memory bandwidth is available only for that node. In Fig.\ref{g1g2}, we show the performance slowdown and increase in memory access cost for all the workloads at bandwidth and latency combination of 100/400 Gbps and 15/5 ns, respectively. As expected, we saw a significant performance slowdown that ranges from 10\% to 120\% for \textbf{Block} compared to \textbf{Local}. The variation in the slowdown for different workloads is due to changing memory access patterns, memory sensitivity, and the number of memory accesses. SC and PF are more sensitive to increased memory latency and face maximum slowdown. HPCG, KD, KC, and PR are less sensitive to memory latency and observe slowdowns between 5\%-18\% even after facing a significant increase in memory access cost. Although most workloads have similar average memory latency with \textbf{Block}, some show more variation when compared to \textbf{Local}. For instance, NW, BT, FT, MG, and FE face the least increase in memory cost, as these workloads already have large memory latency with \textbf{Local} (ranging between 70-100 ns) due to a streaming access pattern as the memory bandwidth is limited. With disaggregated memory, the requests are first distributed across two separate memory units (local and remote), thereby eliminating some contention due to more bandwidth. However, FT, MG, and FE still face a 25\% to 30\% performance slowdown due to a significantly high memory access count, unlike NW and BT, which have fewer memory accesses and did not face much impact from having 50\% of memory footprint in remote memory. MG and PF have large memory accesses, but PF being single-threaded, could not use as much memory parallelism to hide memory latency and face a 2x slowdown.

\begin{figure}[t]
\centerline{\includegraphics[width=14cm,height=2.75cm]{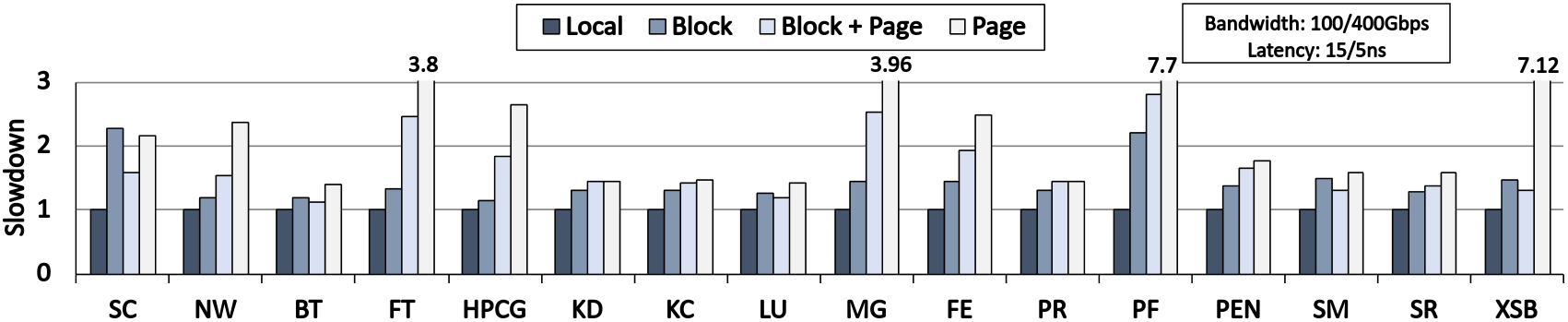}}
\end{figure}
\begin{figure}[t]
\centerline{\includegraphics[width=14cm,height=2.75cm]{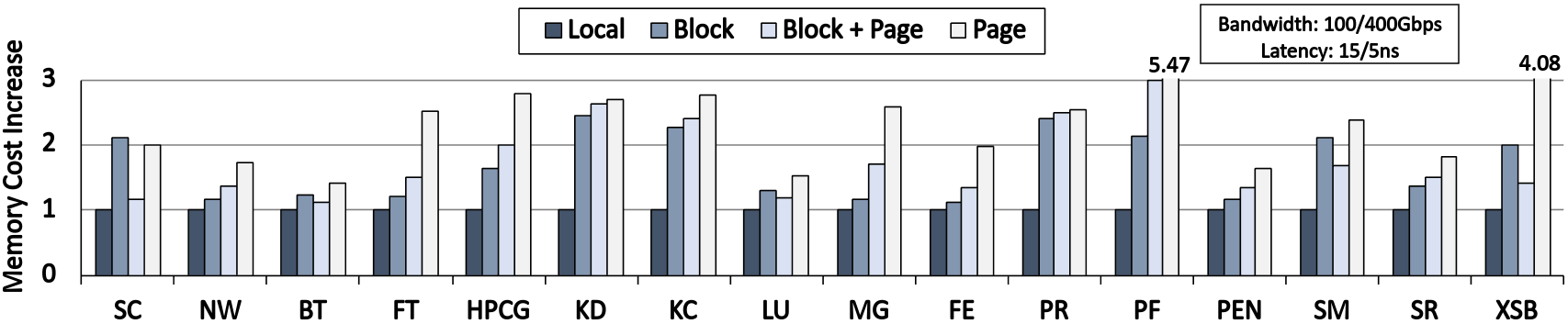}}
\caption{Impact on system performance (top) and memory cost (bottom) on all the workloads over all four configurations}
\label{g1g2}
\end{figure}

With \textbf{Block+Page}, there was an expectation to observe a significant performance improvement compared to baseline \textbf{Block}. However, migrating remote pages to local memory shows no significant benefits. Only five out of all the evaluated workloads: SC, BT, LU, SM, and XSB, could get any benefits, if at all, and faces 5\% to 50\% lesser slowdown than baseline. On the other hand, \textbf{Page} faces the maximum slowdowns compared to all the other scenarios. Firstly, the trend clearly explains the reason behind the performance slowdown with \textbf{Page}, which only accesses remote memory at page granularity and migrates a complete remote page on every LLC miss belonging to a remote address. Remote page accesses are costly as they read multiple cache blocks within a memory page (64 for 4KB page in our case) and add significant transmission delays at the network due to the large packet size (4KB). In the case of \textbf{Block+Page}, only a few hot pages are migrated to local memory and do not see the same performance drop as in \textit{Page}. 
Most of the memory accesses are still performed at block granularity. We observed that SC, BT, and SM all have small memory footprints and fewer memory accesses, so it has only a few pages to migrate. It does not add severe delay to the critical block memory accesses to remote memory while accessing a page. It also allows completing a good percentage of memory accesses in local memory, utilizing the locality in hot pages. For LU and XSB, even though they migrate more pages, more than 90\% of the memory accesses are completed in local memory as the result of migration and hence observe better performance. KD, KC, and PR have a few memory accesses and face a small slowdown due to remote page accesses compared to \textbf{Block}. FT, MG, FE, and PF have enormous memory access traffic and migrate many pages due to their large memory footprints. The performance only gets impacted due to extra delays added by page accesses to regular demand remote memory accesses, which are in their critical path. Similar is the case for HPCG, which is more sensitive to page access delays, despite a small increase in average memory latency. NW and SR also could not benefit from migrating hot pages. Overall, the impact is lesser than \textbf{Page} for all workloads, as the regular block accesses are also performed simultaneously in remote memory. Further, KC, MG, PF, PEN, and SR also show less amount of spatial locality and could only get 20\%, 15\%, 25\%, 23\%, and 18\% additional memory accesses in local memory even after migrating 66\%, 65\%, 48\%, 38\%, and 62\% remote pages to local memory, respectively, out of the 50\% of total pages that are placed in remote.

Page migration has also been widely used in DRM-NVM hybrid memory systems; however, they do not present the same challenges as in disaggregated memory systems. A larger memory footprint on remote memory and the presence of an interconnection network make hot page migration more challenging in disaggregated environments to realize performance gains. We observed a few things that may help improve its performance with page migration. However, we leave it to future research work to explore optimized designs. The continuous access of a complete page for 64 blocks is problematic for two reasons. One is that the pending block memory requests to that page face long delays and are only served once the migration is complete. Two, the block memory requests to other pages in the same remote memory pool may get starved if multiple hot page requests are issued together. Further, optimal scheduling of page migrations becomes more critical due to the same reasons. An optimized hot page migration must consider all these issues for a viable solution. In Daemon \cite{10.1145/3579445}, authors explored bandwidth partitioning, link compression, and schedules page access only when the utilization of the block request queue is less. However, page access is still completed as a whole, adding a delay of around 1.2-1.5µs to subsequent block accesses even with an optimal scheduling policy. With truly disaggregated hardware and CXL interconnects, it is better to break down the page accesses into multiple cache line requests that can reduce most of the overhead with extra architectural support. 

Next, we show the performance impact over different network configurations described earlier. We reduced the node's interface bandwidth by 1/2 to 50Gbps and the switch bandwidth by 1/4 to 100Gbps. However, we kept the same latency at every point. The first graph in Fig. \ref{g3g4g5} shows that \textbf{Block+Page} and \textbf{Page} are significantly impacted by reducing the bandwidth. As both systems access remote memory pages with a response packet size of 4KB, therefore face an additional transmission delays of around 575ns per page during a response by the memory pool. However, workloads show different sensitivity to the increased page access time. On the other hand, the performance of \textbf{Block} is more or less the same as in the previous network configuration, as it only accesses remote memory at 64B block granularity. Although each memory access takes a few nano-seconds more, the continuous memory accesses hide some of these latencies and have no significant drop in performance. 

Next, we increase the latency parameter by 2x and use the same bandwidth of 100/400Gbps. The trend is the opposite in this case, as the packets with small sizes (block access) are most impacted and face almost two times remote memory latency for each access compared to initial network parameters. In contrast, page response packets get a slight increase in latency compared to their overall access time. \textbf{Page} is least impacted by increasing the latency, as each page access adds only 70ns extra for the request and response combined. The performance of \textbf{Block+Page} is also comparatively better in those cases where the number of remote memory accesses was significantly reduced due to page migration. However, it still has to send many block accesses to remote memory, which increases the wait times. In the last case, we decrease both the latency and bandwidth, for which the performance can be seen in the bottom graph in Fig. \ref{g3g4g5}. Overall, in all these cases, BT, KD, KC, LU, PR, and SR are least impacted by the memory disaggregation.

\subsection{Scalable Disaggregated Memory Systems}
As multiple memory pools are expected to be grouped with several compute nodes in a practical disaggregated environment, scalability remains a crucial aspect of disaggregated memory systems. Hence, we evaluate the performance impact of disaggregation in different compute node to memory pool configurations (xC:yM). We kept the memory footprint at a same ratio of 50:50 between local and remote in all the scalable configurations. The remote chunk allocation is done through round-robin pool selection whenever more than one memory pools are there.

\begin{figure}[t]
\centerline{\includegraphics[width=14cm,height=2.75cm]{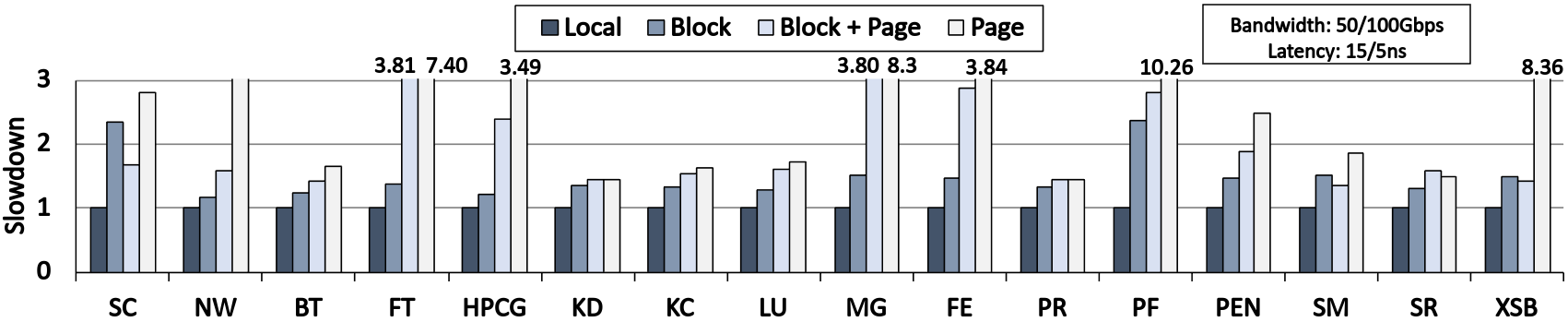}}
\end{figure}
\begin{figure}[t]
\centerline{\includegraphics[width=14cm,height=2.75cm]{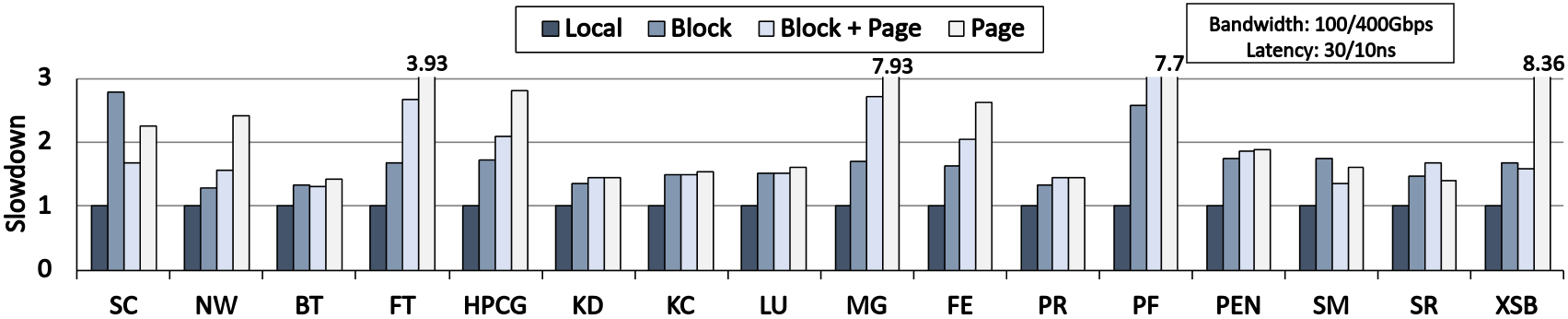}}
\end{figure}
\begin{figure}[t]
\centerline{\includegraphics[width=14cm,height=2.75cm]{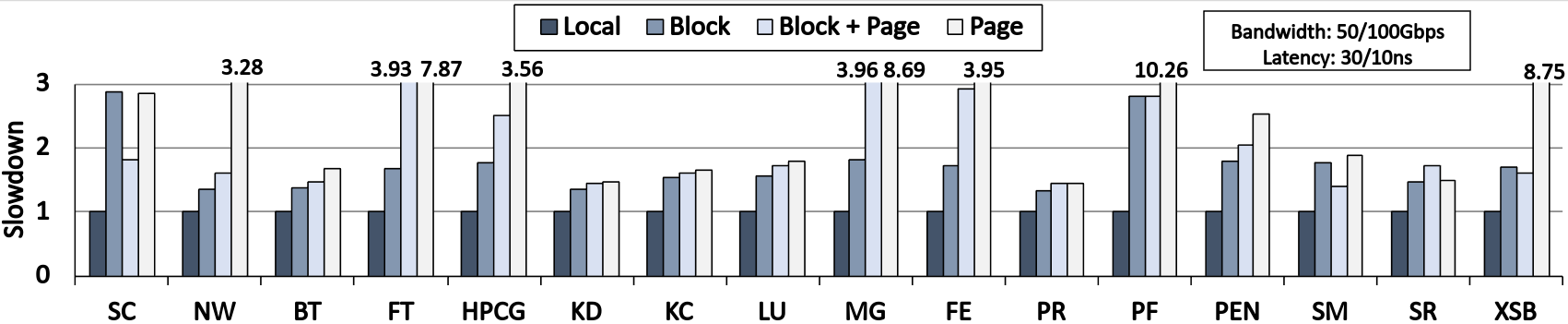}}
\caption{Impact on system performance for all workloads on changing the network latency and bandwidth parameters}
\label{g3g4g5}
\end{figure}

\subsubsection{Sensitivity to Multiple Memory Pools}
Fig. \ref{g6g7} (Top) shows the performance improvement of each workload when a single node is configured with multiple memory pools. The performance is shown for the baseline configuration \textbf{Block}. With 2-memory pools, we observe a maximum performance improvement of around 25\% for SC and FE, while it varies from 2\% to 15\% for most of the workloads and none for NW and SR. The improvement can relate to the increased memory bandwidth with more memory pools, as the remote memory accesses are now distributed across multiple memory units. Thus reducing the chances of contention in the queue and tail latency compared to a single memory pool. However, a further increase in memory pools from 2 to 4 has no additional benefits, as two memory pools could fulfill the maximum bandwidth requirements for all the workloads.

\begin{figure}[t]
\centerline{\includegraphics[width=14cm,height=2.75cm]{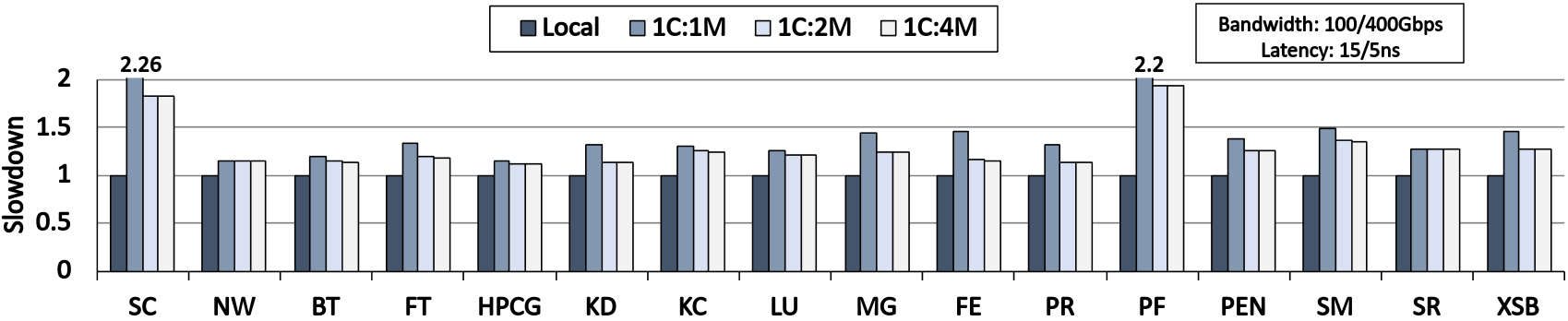}}
\end{figure}
\begin{figure}[t]
\centerline{\includegraphics[width=14cm,height=2.75cm]{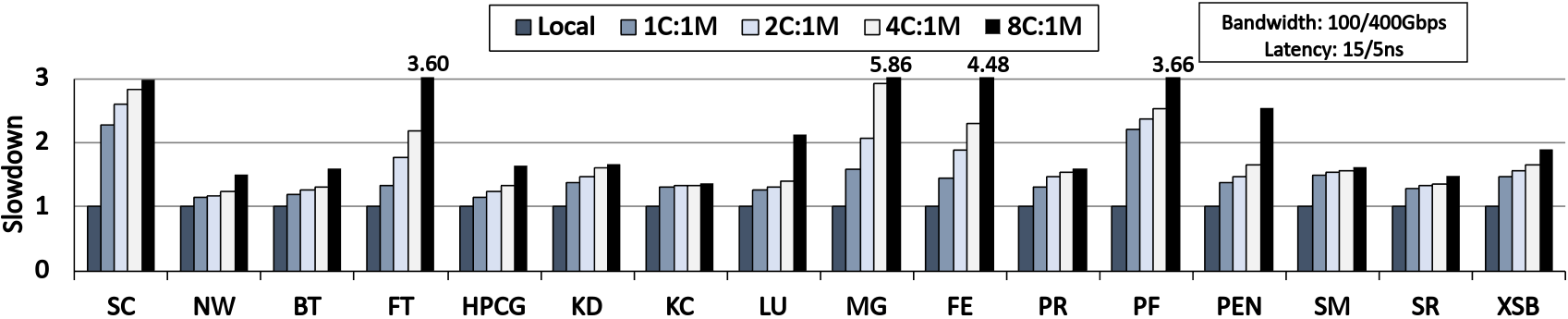}}
\caption{Impact on system performance for all the workloads on increasing the number of memory pools with a single compute pool (Top) or by increasing the number of compute nodes with a single memory pool (Bottom)}
\label{g6g7}
\end{figure}

\begin{figure}[t]
\centerline{\includegraphics[width=14cm,height=2.75cm]{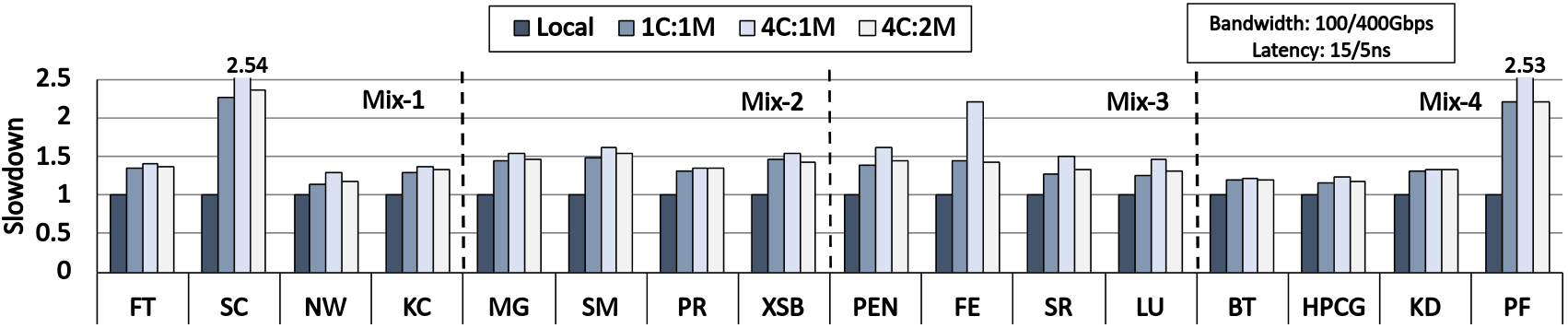}}
\end{figure}
\begin{figure}[t]
\centerline{\includegraphics[width=14cm,height=2.75cm]{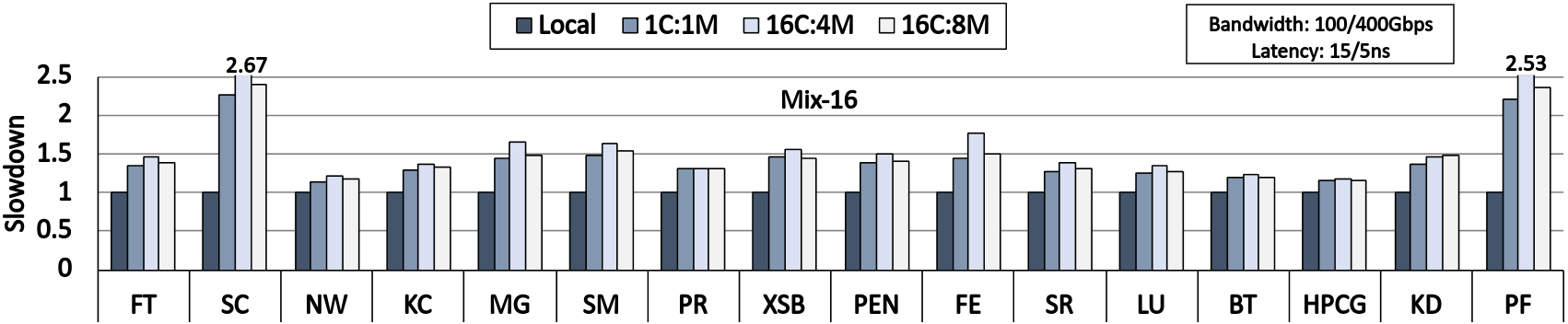}}
\caption{Impact on system performance with different compute-to-memory node configurations and workload combinations over 4 Compute Nodes (Top) and 16 Compute Nodes (Bottom)}
\label{g7g8}
\end{figure}

\subsubsection{Sensitivity to Multiple Compute Nodes}
In Fig. \ref{g6g7} (Bottom), we show the impact of increasing the number of nodes running the same workload with a single memory pool. SC being more memory sensitive is significantly impacted in all configurations with more than one compute node. FT, MG, FE, and PF, due to their extensive memory requests, observe high contention in the memory queues. However, with a 100Gbps interface and 400Gbps network bandwidth, and large enough buffers, interconnect could still manage the high network traffic without much extra overhead. Finally, LU and PEN observed a server slowdown when the nodes were increased from 4 to 8.

\subsubsection{Sensitivity to Multiple Compute and Memory Nodes}
Next, we increase the number of compute nodes and memory pools to evaluate the performance for expected configurations in a practical environment. Fig. \ref{g7g8} (Top) shows the performance impact when four nodes run simultaneously over 1 or 2 memory pools. The performance is shown for four different workload mixes, with one workload running at each node. The workload mixes were created so that the nodes have enough variation in their memory access rates. As expected, we observe a slowdown of 2\% (PR, KC, KD) to 35\% (FE) when 4-compute nodes are configured with only one memory node (4c:1M) compared to a single compute and memory node (1C:1M). Subsequently, this slowdown is 10\% to 125\% compared to a system using only local memory. However, the performance difference is negligible for 4C:2M compared to 1C:4M, making the case strong for compute-to-memory ratio 2:1 (under the given memory bandwidth parameters). We even observe a performance improvement for XSB in 4C:2M compared to 1C:1M, as the distribution of memory accesses suits its access patterns.

Similarly, we run all 16 workloads with an even bigger configuration by deploying them on 16 compute nodes (one workload on each node) with 4 or 8 memory nodes. Fig. \ref{g7g8} (bottom) shows the performance impact for all the workloads in these configurations. We observed that the performance is more or less the same compared to 4 node configurations over similar compute-to-memory node ratios (2:1), with a slight difference for SC and FE.

\begin{figure}[t]
\centerline{\includegraphics[width=14cm,height=2.75cm]{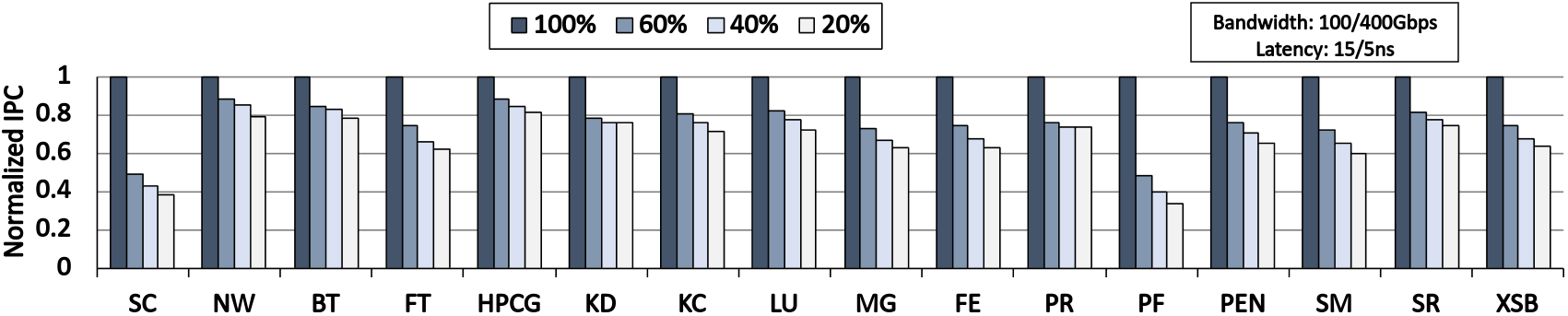}}
\end{figure}
\begin{figure}[t]
\centerline{\includegraphics[width=14cm,height=2.75cm]{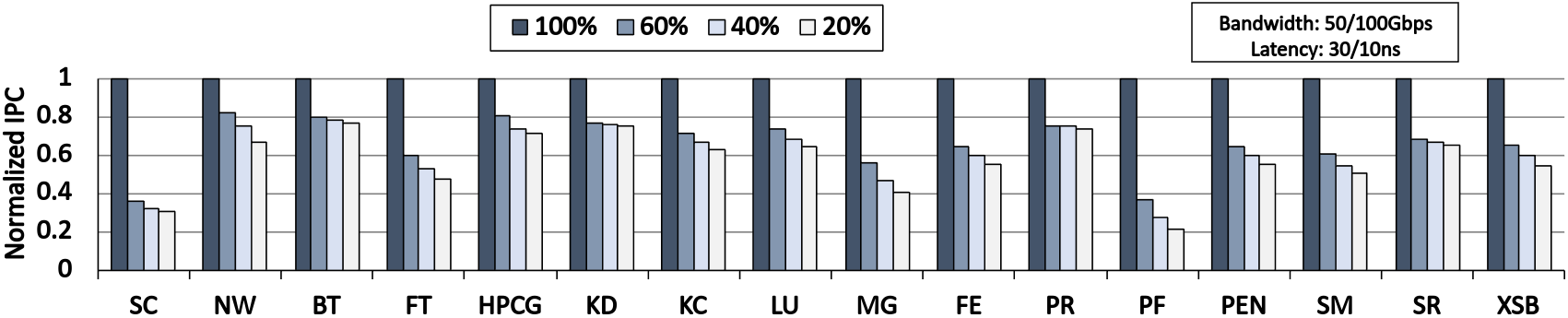}}
\caption{Impact on system performance on changing the local memory footprint}
\label{g9g10}
\end{figure}

\subsection{Sensitivity to Local Memory Footprint}
In a practical setting, the page allocation ratio in local and remote will be crucial in deciding the impact of disaggregation, which varies for different applications, as seen in the above experiments. Therefore we perform a sensitivity analysis on all the workloads with the changing local-to-remote memory footprint ratio. Fig. \ref{g9g10} shows the system IPC at two different network configurations with 60\%, 40\%, and 20\% local memory footprint normalized against 100\% local memory. The findings motivate the requirement of such algorithms, which can decide the local/remote ratio if multiple applications are running on the same node.

\subsection{Network Latency and Bandwidth}
Next, we rigorously test the interconnect model for its correctness using the STREAM benchmark with an input stream array size of 50 million. The simulation is also run for 400 million instructions on the multi-threaded region and has a memory footprint of around 1.14GB. We start with the interface and switch bandwidth of 100 and 400 Gbps, respectively and reduce until up to one-eighth. Similarly, the initial latency is assumed as 15ns (nodes) and 5ns (switch) and increased up to 8 times. Fig. \ref{g11g12} shows the change in IPC and average memory access delay over a range of network bandwidth and latency parameters, respectively.

\begin{figure}[htbp]
\centerline{\includegraphics[width=14cm,height=2.5cm]{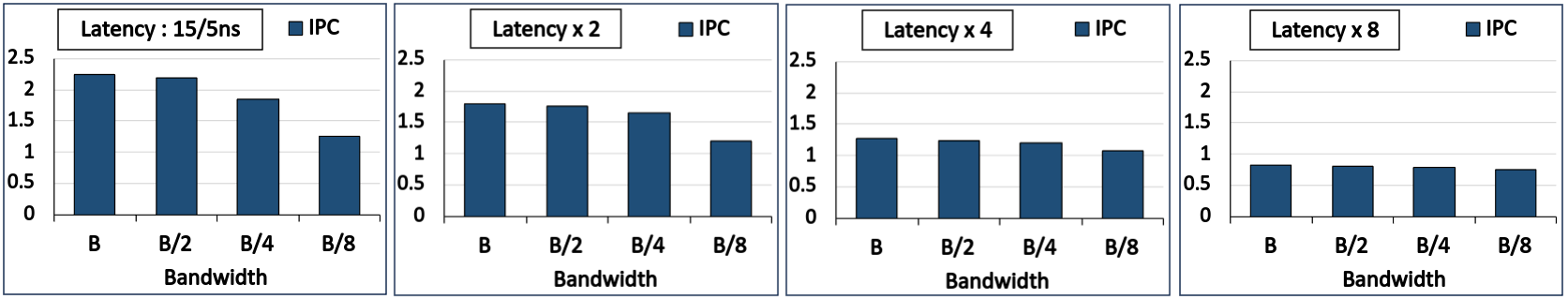}}
\end{figure}
\begin{figure}[htbp]
\centerline{\includegraphics[width=14cm,height=2.5cm]{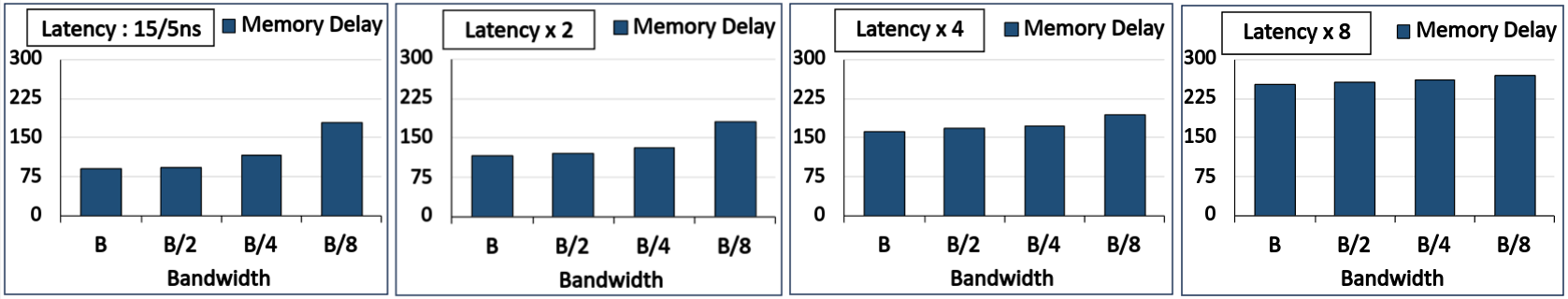}}
\caption{IPC (Top) and Average memory access delay (Bottom) for STREAM benchmark on changing the network bandwidth and latency}
\label{g11g12}
\end{figure}

\section{Conclusion and Future Work}
Disaggregated memory systems are being widely studied for their use in data centers due to their significant advantages over traditional server systems, such as improved memory utilization, scalability, and decoupling of memory. These systems rely on memory-centric interconnects such as CXL to support coherent access to remote memory at block granularity. However, the unavailability of such interconnects hinders path-breaking research ideas from the systems research community. Therefore, we propose a simulator \DESIGN{} that models highly scalable disaggregated memory systems and supports a wide range of user-defined configurations. We perform rigorous validation of \DESIGN{} against gem5 and conduct a broad set of experiments to demonstrate the working of our proposed simulator. We have already made the code \footnote{https://github.com/Amit-P89/-DRackSim} publicly available for the community, and we plan to provide a complete guide for its functionality and code flow once the paper is accepted.

In future work, we plan to add support for shared memory organization and new memory technologies such as DDR5 and HBM. Further, studies are underway to support disaggregated accelerators in data centers \cite{8814493,9820659}. Hence we also plan to extend our tool by integrating an accelerator simulator to move toward a complete simulation infrastructure for disaggregated systems.

\bibliographystyle{ACM-Reference-Format}
\bibliography{ref}


\begin{thebibliography}{39}


\ifx \showCODEN    \undefined \def \showCODEN     #1{\unskip}     \fi
\ifx \showDOI      \undefined \def \showDOI       #1{#1}\fi
\ifx \showISBNx    \undefined \def \showISBNx     #1{\unskip}     \fi
\ifx \showISBNxiii \undefined \def \showISBNxiii  #1{\unskip}     \fi
\ifx \showISSN     \undefined \def \showISSN      #1{\unskip}     \fi
\ifx \showLCCN     \undefined \def \showLCCN      #1{\unskip}     \fi
\ifx \shownote     \undefined \def \shownote      #1{#1}          \fi
\ifx \showarticletitle \undefined \def \showarticletitle #1{#1}   \fi
\ifx \showURL      \undefined \def \showURL       {\relax}        \fi
\providecommand\bibfield[2]{#2}
\providecommand\bibinfo[2]{#2}
\providecommand\natexlab[1]{#1}
\providecommand\showeprint[2][]{arXiv:#2}

\bibitem[Ahn et~al\mbox{.}(2013)]%
        {6557148}
\bibfield{author}{\bibinfo{person}{Jung~Ho Ahn}, \bibinfo{person}{Sheng Li},
  \bibinfo{person}{Seongil O}, {and} \bibinfo{person}{Norman~P. Jouppi}.}
  \bibinfo{year}{2013}\natexlab{}.
\newblock \showarticletitle{McSimA+: A manycore simulator with
  application-level+ simulation and detailed microarchitecture modeling}. In
  \bibinfo{booktitle}{\emph{2013 IEEE International Symposium on Performance
  Analysis of Systems and Software (ISPASS)}}. \bibinfo{publisher}{IEEE
  Computer Society}, \bibinfo{pages}{74--85}.
\newblock
\urldef\tempurl%
\url{https://doi.org/10.1109/ISPASS.2013.6557148}
\showDOI{\tempurl}


\bibitem[Akram and Sawalha(2019)]%
        {8718630}
\bibfield{author}{\bibinfo{person}{Ayaz Akram} {and} \bibinfo{person}{Lina
  Sawalha}.} \bibinfo{year}{2019}\natexlab{}.
\newblock \showarticletitle{A Survey of Computer Architecture Simulation
  Techniques and Tools}.
\newblock \bibinfo{journal}{\emph{IEEE Access}}  \bibinfo{volume}{7}
  (\bibinfo{year}{2019}), \bibinfo{pages}{78120--78145}.
\newblock
\urldef\tempurl%
\url{https://doi.org/10.1109/ACCESS.2019.2917698}
\showDOI{\tempurl}


\bibitem[Al~Maruf and Chowdhury(2020)]%
        {10.5555/3489146.3489204}
\bibfield{author}{\bibinfo{person}{Hasan Al~Maruf} {and}
  \bibinfo{person}{Mosharaf Chowdhury}.} \bibinfo{year}{2020}\natexlab{}.
\newblock \showarticletitle{Effectively Prefetching Remote Memory with Leap}.
  In \bibinfo{booktitle}{\emph{Proceedings of the 2020 USENIX Conference on
  Usenix Annual Technical Conference}} \emph{(\bibinfo{series}{USENIX
  ATC'20})}. \bibinfo{publisher}{USENIX Association}, \bibinfo{address}{USA},
  Article \bibinfo{articleno}{58}, \bibinfo{numpages}{15}~pages.
\newblock
\showISBNx{978-1-939133-14-4}


\bibitem[Bielski et~al\mbox{.}(2018)]%
        {8342174}
\bibfield{author}{\bibinfo{person}{M. Bielski}, \bibinfo{person}{I. Syrigos},
  \bibinfo{person}{K. Katrinis}, \bibinfo{person}{D. Syrivelis},
  \bibinfo{person}{A. Reale}, \bibinfo{person}{D. Theodoropoulos},
  \bibinfo{person}{N. Alachiotis}, \bibinfo{person}{D. Pnevmatikatos},
  \bibinfo{person}{E.H. Pap}, \bibinfo{person}{G. Zervas}, \bibinfo{person}{V.
  Mishra}, \bibinfo{person}{A. Saljoghei}, \bibinfo{person}{A. Rigo},
  \bibinfo{person}{J.~Fernando Zazo}, \bibinfo{person}{S. Lopez-Buedo},
  \bibinfo{person}{M. Torrents}, \bibinfo{person}{F. Zyulkyarov},
  \bibinfo{person}{M. Enrico}, {and} \bibinfo{person}{O.~Gonzalez de Dios}.}
  \bibinfo{year}{2018}\natexlab{}.
\newblock \showarticletitle{dReDBox: Materializing a full-stack rack-scale
  system prototype of a next-generation disaggregated datacenter}. In
  \bibinfo{booktitle}{\emph{2018 Design, Automation \& Test in Europe
  Conference \& Exhibition (DATE)}}. \bibinfo{pages}{1093--1098}.
\newblock
\urldef\tempurl%
\url{https://doi.org/10.23919/DATE.2018.8342174}
\showDOI{\tempurl}


\bibitem[Chang et~al\mbox{.}(2016)]%
        {7753261}
\bibfield{author}{\bibinfo{person}{Yisong Chang}, \bibinfo{person}{Ke Zhang},
  \bibinfo{person}{Sally~A. McKee}, \bibinfo{person}{Lixin Zhang},
  \bibinfo{person}{Mingyu Chen}, \bibinfo{person}{Liqiang Ren}, {and}
  \bibinfo{person}{Zhiwei Xu}.} \bibinfo{year}{2016}\natexlab{}.
\newblock \showarticletitle{Extending On-chip Interconnects for rack-level
  remote resource access}. In \bibinfo{booktitle}{\emph{2016 IEEE 34th
  International Conference on Computer Design (ICCD)}}.
  \bibinfo{pages}{56--63}.
\newblock
\urldef\tempurl%
\url{https://doi.org/10.1109/ICCD.2016.7753261}
\showDOI{\tempurl}


\bibitem[Che et~al\mbox{.}(2009)]%
        {5306797}
\bibfield{author}{\bibinfo{person}{Shuai Che}, \bibinfo{person}{Michael Boyer},
  \bibinfo{person}{Jiayuan Meng}, \bibinfo{person}{David Tarjan},
  \bibinfo{person}{Jeremy~W. Sheaffer}, \bibinfo{person}{Sang-Ha Lee}, {and}
  \bibinfo{person}{Kevin Skadron}.} \bibinfo{year}{2009}\natexlab{}.
\newblock \showarticletitle{Rodinia: A benchmark suite for heterogeneous
  computing}. In \bibinfo{booktitle}{\emph{2009 IEEE International Symposium on
  Workload Characterization (IISWC)}}. \bibinfo{pages}{44--54}.
\newblock
\urldef\tempurl%
\url{https://doi.org/10.1109/IISWC.2009.5306797}
\showDOI{\tempurl}


\bibitem[Crozier et~al\mbox{.}(2009)]%
        {osti_993908}
\bibfield{author}{\bibinfo{person}{Paul~Stewart Crozier},
  \bibinfo{person}{Heidi~K Thornquist}, \bibinfo{person}{Robert~W Numrich},
  \bibinfo{person}{Alan~B Williams}, \bibinfo{person}{Harold~Carter Edwards},
  \bibinfo{person}{Eric~Richard Keiter}, \bibinfo{person}{Mahesh Rajan},
  \bibinfo{person}{James~M Willenbring}, \bibinfo{person}{Douglas~W Doerfler},
  {and} \bibinfo{person}{Michael~Allen Heroux}.}
  \bibinfo{year}{2009}\natexlab{}.
\newblock \showarticletitle{Improving performance via mini-applications.}
\newblock  (\bibinfo{date}{9} \bibinfo{year}{2009}).
\newblock
\urldef\tempurl%
\url{https://doi.org/10.2172/993908}
\showDOI{\tempurl}


\bibitem[Dragojevi\'{c} et~al\mbox{.}(2014)]%
        {10.5555/2616448.2616486}
\bibfield{author}{\bibinfo{person}{Aleksandar Dragojevi\'{c}},
  \bibinfo{person}{Dushyanth Narayanan}, \bibinfo{person}{Orion Hodson}, {and}
  \bibinfo{person}{Miguel Castro}.} \bibinfo{year}{2014}\natexlab{}.
\newblock \showarticletitle{FaRM: Fast Remote Memory}. In
  \bibinfo{booktitle}{\emph{Proceedings of the 11th USENIX Conference on
  Networked Systems Design and Implementation}} (Seattle, WA)
  \emph{(\bibinfo{series}{NSDI'14})}. \bibinfo{publisher}{USENIX Association},
  \bibinfo{address}{USA}, \bibinfo{pages}{401–414}.
\newblock
\showISBNx{9781931971096}


\bibitem[Express({[n.\,d.]})]%
        {Express}
\bibfield{author}{\bibinfo{person}{CXL Express}.}
  \bibinfo{year}{[n.\,d.]}\natexlab{}.
\newblock \bibinfo{booktitle}{\emph{CXL specification}}.
\newblock
\urldef\tempurl%
\url{https://www.computeexpresslink.org/download-the-specification}
\showURL{%
\tempurl}


\bibitem[Ferenbaugh(2015)]%
        {https://doi.org/10.1002/cpe.3422}
\bibfield{author}{\bibinfo{person}{Charles~R. Ferenbaugh}.}
  \bibinfo{year}{2015}\natexlab{}.
\newblock \showarticletitle{PENNANT: an unstructured mesh mini-app for advanced
  architecture research}.
\newblock \bibinfo{journal}{\emph{Concurrency and Computation: Practice and
  Experience}} \bibinfo{volume}{27}, \bibinfo{number}{17}
  (\bibinfo{year}{2015}), \bibinfo{pages}{4555--4572}.
\newblock
\urldef\tempurl%
\url{https://doi.org/10.1002/cpe.3422}
\showDOI{\tempurl}
\showeprint{https://onlinelibrary.wiley.com/doi/pdf/10.1002/cpe.3422}


\bibitem[Fingler et~al\mbox{.}(2022)]%
        {9820659}
\bibfield{author}{\bibinfo{person}{Henrique Fingler}, \bibinfo{person}{Zhiting
  Zhu}, \bibinfo{person}{Esther Yoon}, \bibinfo{person}{Zhipeng Jia},
  \bibinfo{person}{Emmett Witchel}, {and} \bibinfo{person}{Christopher~J.
  Rossbach}.} \bibinfo{year}{2022}\natexlab{}.
\newblock \showarticletitle{DGSF: Disaggregated GPUs for Serverless Functions}.
  In \bibinfo{booktitle}{\emph{2022 IEEE International Parallel and Distributed
  Processing Symposium (IPDPS)}}. \bibinfo{pages}{739--750}.
\newblock
\urldef\tempurl%
\url{https://doi.org/10.1109/IPDPS53621.2022.00077}
\showDOI{\tempurl}


\bibitem[Gao et~al\mbox{.}(2016)]%
        {10.5555/3026877.3026897}
\bibfield{author}{\bibinfo{person}{Peter~X. Gao}, \bibinfo{person}{Akshay
  Narayan}, \bibinfo{person}{Sagar Karandikar}, \bibinfo{person}{Joao
  Carreira}, \bibinfo{person}{Sangjin Han}, \bibinfo{person}{Rachit Agarwal},
  \bibinfo{person}{Sylvia Ratnasamy}, {and} \bibinfo{person}{Scott Shenker}.}
  \bibinfo{year}{2016}\natexlab{}.
\newblock \showarticletitle{Network Requirements for Resource Disaggregation}.
  In \bibinfo{booktitle}{\emph{Proceedings of the 12th USENIX Conference on
  Operating Systems Design and Implementation}} (Savannah, GA, USA)
  \emph{(\bibinfo{series}{OSDI'16})}. \bibinfo{publisher}{USENIX Association},
  \bibinfo{address}{USA}, \bibinfo{pages}{249–264}.
\newblock
\showISBNx{9781931971331}


\bibitem[Giannoula et~al\mbox{.}(2023)]%
        {10.1145/3579445}
\bibfield{author}{\bibinfo{person}{Christina Giannoula},
  \bibinfo{person}{Kailong Huang}, \bibinfo{person}{Jonathan Tang},
  \bibinfo{person}{Nectarios Koziris}, \bibinfo{person}{Georgios Goumas},
  \bibinfo{person}{Zeshan Chishti}, {and} \bibinfo{person}{Nandita
  Vijaykumar}.} \bibinfo{year}{2023}\natexlab{}.
\newblock \showarticletitle{DaeMon: Architectural Support for Efficient Data
  Movement in Fully Disaggregated Systems}.
\newblock \bibinfo{journal}{\emph{Proc. ACM Meas. Anal. Comput. Syst.}}
  \bibinfo{volume}{7}, \bibinfo{number}{1}, Article \bibinfo{articleno}{16}
  (\bibinfo{date}{mar} \bibinfo{year}{2023}), \bibinfo{numpages}{36}~pages.
\newblock
\urldef\tempurl%
\url{https://doi.org/10.1145/3579445}
\showDOI{\tempurl}


\bibitem[Griebler et~al\mbox{.}(2018)]%
        {8374543}
\bibfield{author}{\bibinfo{person}{Dalvan Griebler}, \bibinfo{person}{Junior
  Loff}, \bibinfo{person}{Gabriele Mencagli}, \bibinfo{person}{Marco
  Danelutto}, {and} \bibinfo{person}{Luiz~Gustavo Fernandes}.}
  \bibinfo{year}{2018}\natexlab{}.
\newblock \showarticletitle{Efficient NAS Benchmark Kernels with C++ Parallel
  Programming}. In \bibinfo{booktitle}{\emph{2018 26th Euromicro International
  Conference on Parallel, Distributed and Network-based Processing (PDP)}}.
  \bibinfo{pages}{733--740}.
\newblock
\urldef\tempurl%
\url{https://doi.org/10.1109/PDP2018.2018.00120}
\showDOI{\tempurl}


\bibitem[Gu et~al\mbox{.}(2017)]%
        {10.5555/3154630.3154683}
\bibfield{author}{\bibinfo{person}{Juncheng Gu}, \bibinfo{person}{Youngmoon
  Lee}, \bibinfo{person}{Yiwen Zhang}, \bibinfo{person}{Mosharaf Chowdhury},
  {and} \bibinfo{person}{Kang~G. Shin}.} \bibinfo{year}{2017}\natexlab{}.
\newblock \showarticletitle{Efficient Memory Disaggregation with INFINISWAP}.
  In \bibinfo{booktitle}{\emph{Proceedings of the 14th USENIX Conference on
  Networked Systems Design and Implementation}} (Boston, MA, USA)
  \emph{(\bibinfo{series}{NSDI'17})}. \bibinfo{publisher}{USENIX Association},
  \bibinfo{address}{USA}, \bibinfo{pages}{649–667}.
\newblock
\showISBNx{9781931971379}


\bibitem[Guleria et~al\mbox{.}(2019)]%
        {8814493}
\bibfield{author}{\bibinfo{person}{Anubhav Guleria}, \bibinfo{person}{J
  Lakshmi}, {and} \bibinfo{person}{Chakri Padala}.}
  \bibinfo{year}{2019}\natexlab{}.
\newblock \showarticletitle{QuADD: QUantifying Accelerator Disaggregated
  Datacenter Efficiency}. In \bibinfo{booktitle}{\emph{2019 IEEE 12th
  International Conference on Cloud Computing (CLOUD)}}.
  \bibinfo{pages}{349--357}.
\newblock
\urldef\tempurl%
\url{https://doi.org/10.1109/CLOUD.2019.00064}
\showDOI{\tempurl}


\bibitem[Gunow et~al\mbox{.}(2015)]%
        {Gunow2015}
\bibfield{author}{\bibinfo{person}{Geoffrey Gunow}, \bibinfo{person}{John
  Tramm}, \bibinfo{person}{Benoit Forget}, \bibinfo{person}{Kord Smith}, {and}
  \bibinfo{person}{Tim He}.} \bibinfo{year}{2015}\natexlab{}.
\newblock \showarticletitle{{SimpleMOC} -- A PERFORMANCE ABSTRACTION FOR {3D
  MOC}}. In \bibinfo{booktitle}{\emph{ANS \& M\&C 2015 - Joint International
  Conference on Mathematics and Computation (M\&C), Supercomputing in Nuclear
  Applications (SNA) and the Monte Carlo (MC) Method}}.
\newblock


\bibitem[Hong et~al\mbox{.}(2020)]%
        {9137193}
\bibfield{author}{\bibinfo{person}{Seokbin Hong}, \bibinfo{person}{Won-Ok
  Kwon}, {and} \bibinfo{person}{Myeong-Hoon Oh}.}
  \bibinfo{year}{2020}\natexlab{}.
\newblock \showarticletitle{Hardware Implementation and Analysis of Gen-Z
  Protocol for Memory-Centric Architecture}.
\newblock \bibinfo{journal}{\emph{IEEE Access}}  \bibinfo{volume}{8}
  (\bibinfo{year}{2020}), \bibinfo{pages}{127244--127253}.
\newblock
\urldef\tempurl%
\url{https://doi.org/10.1109/ACCESS.2020.3008227}
\showDOI{\tempurl}


\bibitem[HPCG(2019)]%
        {hpcg}
\bibfield{author}{\bibinfo{person}{HPCG}.} \bibinfo{year}{2019}\natexlab{}.
\newblock \bibinfo{title}{{G}it{H}ub - hpcg-benchmark/hpcg: {O}fficial
  {H}{P}{C}{G} benchmark source code --- github.com}.
\newblock
  \bibinfo{howpublished}{\url{https://github.com/hpcg-benchmark/hpcg/}}.
\newblock
\newblock
\shownote{[Accessed 18-Jul-2023]}.


\bibitem[Intel({[n.\,d.]})]%
        {intelIntelRack}
\bibfield{author}{\bibinfo{person}{Intel}.}
  \bibinfo{year}{[n.\,d.]}\natexlab{}.
\newblock \bibinfo{title}{{I}ntel® {R}ack {S}cale {D}esign ({I}ntel®
  {R}{S}{D}) {A}rchitecture {W}hite {P}aper --- intel.in}.
\newblock
  \bibinfo{howpublished}{\url{https://www.intel.in/content/www/in/en/architecture-and-technology/rack-scale-design/rack-scale-design-architecture-white-paper.html}}.
\newblock
\newblock
\shownote{[Accessed 02-Jul-2023]}.


\bibitem[Islam et~al\mbox{.}(2020)]%
        {10.1145/3364179}
\bibfield{author}{\bibinfo{person}{Mahzabeen Islam}, \bibinfo{person}{Shashank
  Adavally}, \bibinfo{person}{Marko Scrbak}, {and} \bibinfo{person}{Krishna
  Kavi}.} \bibinfo{year}{2020}\natexlab{}.
\newblock \showarticletitle{On-the-Fly Page Migration and Address
  Reconciliation for Heterogeneous Memory Systems}.
\newblock \bibinfo{journal}{\emph{J. Emerg. Technol. Comput. Syst.}}
  \bibinfo{volume}{16}, \bibinfo{number}{1}, Article \bibinfo{articleno}{10}
  (\bibinfo{date}{jan} \bibinfo{year}{2020}), \bibinfo{numpages}{27}~pages.
\newblock
\showISSN{1550-4832}
\urldef\tempurl%
\url{https://doi.org/10.1145/3364179}
\showDOI{\tempurl}


\bibitem[Karlin et~al\mbox{.}(2013)]%
        {osti_1090032}
\bibfield{author}{\bibinfo{person}{I Karlin}, \bibinfo{person}{J Keasler},
  {and} \bibinfo{person}{J~R Neely}.} \bibinfo{year}{2013}\natexlab{}.
\newblock \showarticletitle{LULESH 2.0 Updates and Changes}.
\newblock  (\bibinfo{date}{7} \bibinfo{year}{2013}).
\newblock
\urldef\tempurl%
\url{https://doi.org/10.2172/1090032}
\showDOI{\tempurl}


\bibitem[Katrinis et~al\mbox{.}(2016)]%
        {7459397}
\bibfield{author}{\bibinfo{person}{K. Katrinis}, \bibinfo{person}{D.
  Syrivelis}, \bibinfo{person}{D. Pnevmatikatos}, \bibinfo{person}{G. Zervas},
  \bibinfo{person}{D. Theodoropoulos}, \bibinfo{person}{I. Koutsopoulos},
  \bibinfo{person}{K. Hasharoni}, \bibinfo{person}{D. Raho},
  \bibinfo{person}{C. Pinto}, \bibinfo{person}{F. Espina}, \bibinfo{person}{S.
  Lopez-Buedo}, \bibinfo{person}{Q. Chen}, \bibinfo{person}{M. Nemirovsky},
  \bibinfo{person}{D. Roca}, \bibinfo{person}{H. Klos}, {and}
  \bibinfo{person}{T. Berends}.} \bibinfo{year}{2016}\natexlab{}.
\newblock \showarticletitle{Rack-scale disaggregated cloud data centers: The
  dReDBox project vision}. In \bibinfo{booktitle}{\emph{2016 Design, Automation
  \& Test in Europe Conference \& Exhibition (DATE)}}.
  \bibinfo{pages}{690--695}.
\newblock


\bibitem[Lee et~al\mbox{.}(2021)]%
        {10.1145/3477132.3483561}
\bibfield{author}{\bibinfo{person}{Seung-seob Lee}, \bibinfo{person}{Yanpeng
  Yu}, \bibinfo{person}{Yupeng Tang}, \bibinfo{person}{Anurag Khandelwal},
  \bibinfo{person}{Lin Zhong}, {and} \bibinfo{person}{Abhishek Bhattacharjee}.}
  \bibinfo{year}{2021}\natexlab{}.
\newblock \showarticletitle{MIND: In-Network Memory Management for
  Disaggregated Data Centers}. In \bibinfo{booktitle}{\emph{Proceedings of the
  ACM SIGOPS 28th Symposium on Operating Systems Principles}} (Virtual Event,
  Germany) \emph{(\bibinfo{series}{SOSP '21})}. \bibinfo{publisher}{Association
  for Computing Machinery}, \bibinfo{address}{New York, NY, USA},
  \bibinfo{pages}{488–504}.
\newblock
\showISBNx{9781450387095}
\urldef\tempurl%
\url{https://doi.org/10.1145/3477132.3483561}
\showDOI{\tempurl}


\bibitem[Lim et~al\mbox{.}(2009)]%
        {10.1145/1555754.1555789}
\bibfield{author}{\bibinfo{person}{Kevin Lim}, \bibinfo{person}{Jichuan Chang},
  \bibinfo{person}{Trevor Mudge}, \bibinfo{person}{Parthasarathy Ranganathan},
  \bibinfo{person}{Steven~K. Reinhardt}, {and} \bibinfo{person}{Thomas~F.
  Wenisch}.} \bibinfo{year}{2009}\natexlab{}.
\newblock \showarticletitle{Disaggregated Memory for Expansion and Sharing in
  Blade Servers}. In \bibinfo{booktitle}{\emph{Proceedings of the 36th Annual
  International Symposium on Computer Architecture}} (Austin, TX, USA)
  \emph{(\bibinfo{series}{ISCA '09})}. \bibinfo{publisher}{Association for
  Computing Machinery}, \bibinfo{address}{New York, NY, USA},
  \bibinfo{pages}{267–278}.
\newblock
\showISBNx{9781605585260}
\urldef\tempurl%
\url{https://doi.org/10.1145/1555754.1555789}
\showDOI{\tempurl}


\bibitem[Lim et~al\mbox{.}(2012)]%
        {6168955}
\bibfield{author}{\bibinfo{person}{Kevin Lim}, \bibinfo{person}{Yoshio Turner},
  \bibinfo{person}{Jose~Renato Santos}, \bibinfo{person}{Alvin AuYoung},
  \bibinfo{person}{Jichuan Chang}, \bibinfo{person}{Parthasarathy Ranganathan},
  {and} \bibinfo{person}{Thomas~F. Wenisch}.} \bibinfo{year}{2012}\natexlab{}.
\newblock \showarticletitle{System-level implications of disaggregated memory}.
  In \bibinfo{booktitle}{\emph{IEEE International Symposium on High-Performance
  Comp Architecture}}. \bibinfo{pages}{1--12}.
\newblock
\urldef\tempurl%
\url{https://doi.org/10.1109/HPCA.2012.6168955}
\showDOI{\tempurl}


\bibitem[Loh et~al\mbox{.}(2009)]%
        {4919638}
\bibfield{author}{\bibinfo{person}{Gabriel~H. Loh}, \bibinfo{person}{Samantika
  Subramaniam}, {and} \bibinfo{person}{Yuejian Xie}.}
  \bibinfo{year}{2009}\natexlab{}.
\newblock \showarticletitle{Zesto: A cycle-level simulator for highly detailed
  microarchitecture exploration}. In \bibinfo{booktitle}{\emph{2009 IEEE
  International Symposium on Performance Analysis of Systems and Software}}.
  \bibinfo{pages}{53--64}.
\newblock
\urldef\tempurl%
\url{https://doi.org/10.1109/ISPASS.2009.4919638}
\showDOI{\tempurl}


\bibitem[Luk et~al\mbox{.}(2005)]%
        {10.1145/1064978.1065034}
\bibfield{author}{\bibinfo{person}{Chi-Keung Luk}, \bibinfo{person}{Robert
  Cohn}, \bibinfo{person}{Robert Muth}, \bibinfo{person}{Harish Patil},
  \bibinfo{person}{Artur Klauser}, \bibinfo{person}{Geoff Lowney},
  \bibinfo{person}{Steven Wallace}, \bibinfo{person}{Vijay~Janapa Reddi}, {and}
  \bibinfo{person}{Kim Hazelwood}.} \bibinfo{year}{2005}\natexlab{}.
\newblock \showarticletitle{Pin: Building Customized Program Analysis Tools
  with Dynamic Instrumentation}.
\newblock \bibinfo{journal}{\emph{SIGPLAN Not.}} \bibinfo{volume}{40},
  \bibinfo{number}{6} (\bibinfo{date}{jun} \bibinfo{year}{2005}),
  \bibinfo{pages}{190–200}.
\newblock
\showISSN{0362-1340}
\urldef\tempurl%
\url{https://doi.org/10.1145/1064978.1065034}
\showDOI{\tempurl}


\bibitem[Novakovic et~al\mbox{.}(2014)]%
        {10.1145/2541940.2541965}
\bibfield{author}{\bibinfo{person}{Stanko Novakovic},
  \bibinfo{person}{Alexandros Daglis}, \bibinfo{person}{Edouard Bugnion},
  \bibinfo{person}{Babak Falsafi}, {and} \bibinfo{person}{Boris Grot}.}
  \bibinfo{year}{2014}\natexlab{}.
\newblock \showarticletitle{Scale-out NUMA}. In
  \bibinfo{booktitle}{\emph{Proceedings of the 19th International Conference on
  Architectural Support for Programming Languages and Operating Systems}} (Salt
  Lake City, Utah, USA) \emph{(\bibinfo{series}{ASPLOS '14})}.
  \bibinfo{publisher}{Association for Computing Machinery},
  \bibinfo{address}{New York, NY, USA}, \bibinfo{pages}{3–18}.
\newblock
\showISBNx{9781450323055}
\urldef\tempurl%
\url{https://doi.org/10.1145/2541940.2541965}
\showDOI{\tempurl}


\bibitem[Pinto et~al\mbox{.}(2020)]%
        {9252003}
\bibfield{author}{\bibinfo{person}{Christian Pinto}, \bibinfo{person}{Dimitris
  Syrivelis}, \bibinfo{person}{Michele Gazzetti}, \bibinfo{person}{Panos
  Koutsovasilis}, \bibinfo{person}{Andrea Reale}, \bibinfo{person}{Kostas
  Katrinis}, {and} \bibinfo{person}{H.~Peter Hofstee}.}
  \bibinfo{year}{2020}\natexlab{}.
\newblock \showarticletitle{ThymesisFlow: A Software-Defined, HW/SW co-Designed
  Interconnect Stack for Rack-Scale Memory Disaggregation}. In
  \bibinfo{booktitle}{\emph{2020 53rd Annual IEEE/ACM International Symposium
  on Microarchitecture (MICRO)}}. \bibinfo{pages}{868--880}.
\newblock
\urldef\tempurl%
\url{https://doi.org/10.1109/MICRO50266.2020.00075}
\showDOI{\tempurl}


\bibitem[Puri et~al\mbox{.}(2022)]%
        {10074901}
\bibfield{author}{\bibinfo{person}{Amit Puri}, \bibinfo{person}{John Jose},
  {and} \bibinfo{person}{Tamarapalli Venkatesh}.}
  \bibinfo{year}{2022}\natexlab{}.
\newblock \showarticletitle{Design and Evaluation of a Rack-Scale Disaggregated
  Memory Architecture For Data Centers}. In \bibinfo{booktitle}{\emph{2022 IEEE
  24th Int Conf on High Performance Computing \& Communications; 8th Int Conf
  on Data Science \& Systems; 20th Int Conf on Smart City; 8th Int Conf on
  Dependability in Sensor, Cloud \& Big Data Systems \& Application
  (HPCC/DSS/SmartCity/DependSys)}}. \bibinfo{pages}{212--217}.
\newblock
\urldef\tempurl%
\url{https://doi.org/10.1109/HPCC-DSS-SmartCity-DependSys57074.2022.00060}
\showDOI{\tempurl}


\bibitem[Quiroga et~al\mbox{.}(2019)]%
        {10.1145/3339985.3358496}
\bibfield{author}{\bibinfo{person}{Josue~V. Quiroga}, \bibinfo{person}{Marti
  Torrents}, \bibinfo{person}{Nehir Sonmez}, \bibinfo{person}{Dimitris
  Theodoropoulos}, \bibinfo{person}{Ferad Zyulkyarov}, {and}
  \bibinfo{person}{Mario Nemirovsky}.} \bibinfo{year}{2019}\natexlab{}.
\newblock \showarticletitle{Evaluation of a Rack-Scale Disaggregated Memory
  Prototype for Cloud Data Centers}. In \bibinfo{booktitle}{\emph{Proceedings
  of the 30th International Workshop on Rapid System Prototyping (RSP'19)}}
  (New York, NY, USA) \emph{(\bibinfo{series}{RSP '19})}.
  \bibinfo{publisher}{Association for Computing Machinery},
  \bibinfo{address}{New York, NY, USA}, \bibinfo{pages}{15–21}.
\newblock
\showISBNx{9781450368476}
\urldef\tempurl%
\url{https://doi.org/10.1145/3339985.3358496}
\showDOI{\tempurl}


\bibitem[Rosenfeld et~al\mbox{.}(2011)]%
        {10.1109/L-CA.2011.4}
\bibfield{author}{\bibinfo{person}{Paul Rosenfeld}, \bibinfo{person}{Elliott
  Cooper-Balis}, {and} \bibinfo{person}{Bruce Jacob}.}
  \bibinfo{year}{2011}\natexlab{}.
\newblock \showarticletitle{DRAMSim2: A Cycle Accurate Memory System
  Simulator}.
\newblock \bibinfo{journal}{\emph{IEEE Comput. Archit. Lett.}}
  \bibinfo{volume}{10}, \bibinfo{number}{1} (\bibinfo{date}{jan}
  \bibinfo{year}{2011}), \bibinfo{pages}{16–19}.
\newblock
\showISSN{1556-6056}
\urldef\tempurl%
\url{https://doi.org/10.1109/L-CA.2011.4}
\showDOI{\tempurl}


\bibitem[Sakalis et~al\mbox{.}(2016)]%
        {7482078}
\bibfield{author}{\bibinfo{person}{Christos Sakalis}, \bibinfo{person}{Carl
  Leonardsson}, \bibinfo{person}{Stefanos Kaxiras}, {and}
  \bibinfo{person}{Alberto Ros}.} \bibinfo{year}{2016}\natexlab{}.
\newblock \showarticletitle{Splash-3: A properly synchronized benchmark suite
  for contemporary research}. In \bibinfo{booktitle}{\emph{2016 IEEE
  International Symposium on Performance Analysis of Systems and Software
  (ISPASS)}}. \bibinfo{pages}{101--111}.
\newblock
\urldef\tempurl%
\url{https://doi.org/10.1109/ISPASS.2016.7482078}
\showDOI{\tempurl}


\bibitem[Sharma(2023)]%
        {9982424}
\bibfield{author}{\bibinfo{person}{Debendra~Das Sharma}.}
  \bibinfo{year}{2023}\natexlab{}.
\newblock \showarticletitle{Compute Express Link (CXL): Enabling Heterogeneous
  Data-Centric Computing With Heterogeneous Memory Hierarchy}.
\newblock \bibinfo{journal}{\emph{IEEE Micro}} \bibinfo{volume}{43},
  \bibinfo{number}{2} (\bibinfo{year}{2023}), \bibinfo{pages}{99--109}.
\newblock
\urldef\tempurl%
\url{https://doi.org/10.1109/MM.2022.3228561}
\showDOI{\tempurl}


\bibitem[Shun and Blelloch(2013)]%
        {10.1145/2442516.2442530}
\bibfield{author}{\bibinfo{person}{Julian Shun} {and} \bibinfo{person}{Guy~E.
  Blelloch}.} \bibinfo{year}{2013}\natexlab{}.
\newblock \showarticletitle{Ligra: A Lightweight Graph Processing Framework for
  Shared Memory}. In \bibinfo{booktitle}{\emph{Proceedings of the 18th ACM
  SIGPLAN Symposium on Principles and Practice of Parallel Programming}}
  (Shenzhen, China) \emph{(\bibinfo{series}{PPoPP '13})}.
  \bibinfo{publisher}{Association for Computing Machinery},
  \bibinfo{address}{New York, NY, USA}, \bibinfo{pages}{135–146}.
\newblock
\showISBNx{9781450319225}
\urldef\tempurl%
\url{https://doi.org/10.1145/2442516.2442530}
\showDOI{\tempurl}


\bibitem[Taylor(2015)]%
        {Taylor:15}
\bibfield{author}{\bibinfo{person}{Jason Taylor}.}
  \bibinfo{year}{2015}\natexlab{}.
\newblock \showarticletitle{Facebook's Data Center Infrastructure: Open
  Compute, Disaggregated Rack, and Beyond}, In \bibinfo{booktitle}{Optical
  Fiber Communication Conference}.
\newblock \bibinfo{journal}{\emph{Optical Fiber Communication Conference}},
  \bibinfo{pages}{W1D.5}.
\newblock
\urldef\tempurl%
\url{https://doi.org/10.1364/OFC.2015.W1D.5}
\showDOI{\tempurl}


\bibitem[Tramm et~al\mbox{.}(2014)]%
        {Tramm:wy}
\bibfield{author}{\bibinfo{person}{John~R Tramm}, \bibinfo{person}{Andrew~R
  Siegel}, \bibinfo{person}{Tanzima Islam}, {and} \bibinfo{person}{Martin
  Schulz}.} \bibinfo{year}{2014}\natexlab{}.
\newblock \showarticletitle{{XSBench} - The Development and Verification of a
  Performance Abstraction for {M}onte {C}arlo Reactor Analysis}. In
  \bibinfo{booktitle}{\emph{{PHYSOR} 2014 - The Role of Reactor Physics toward
  a Sustainable Future}}. \bibinfo{address}{Kyoto}.
\newblock
\urldef\tempurl%
\url{https://www.mcs.anl.gov/papers/P5064-0114.pdf}
\showURL{%
\tempurl}


\bibitem[Zhao et~al\mbox{.}(2019)]%
        {10.1145/3310360}
\bibfield{author}{\bibinfo{person}{Boyan Zhao}, \bibinfo{person}{Rui Hou},
  \bibinfo{person}{Jianbo Dong}, \bibinfo{person}{Michael Huang},
  \bibinfo{person}{Sally~A. Mckee}, \bibinfo{person}{Qianlong Zhang},
  \bibinfo{person}{Yueji Liu}, \bibinfo{person}{Ye Li}, \bibinfo{person}{Lixin
  Zhang}, {and} \bibinfo{person}{Dan Meng}.} \bibinfo{year}{2019}\natexlab{}.
\newblock \showarticletitle{Venice: An Effective Resource Sharing Architecture
  for Data Center Servers}.
\newblock \bibinfo{journal}{\emph{ACM Trans. Comput. Syst.}}
  \bibinfo{volume}{36}, \bibinfo{number}{1}, Article \bibinfo{articleno}{2}
  (\bibinfo{date}{mar} \bibinfo{year}{2019}), \bibinfo{numpages}{26}~pages.
\newblock
\showISSN{0734-2071}
\urldef\tempurl%
\url{https://doi.org/10.1145/3310360}
\showDOI{\tempurl}


\end{thebibliography}

\appendix

\end{document}